\documentclass[final,3p,times,square,comma]{elsarticle}
\pdfoutput=1
\usepackage[switch]{lineno}
\usepackage[utf8]{inputenc}
\usepackage{xcolor}
\usepackage{tabularx,booktabs}
\usepackage{array}
\usepackage{graphicx}
\usepackage{tikz}
\usetikzlibrary{shapes.geometric, arrows}
\usepackage{pgfplots}
\usepackage{xfrac}
\pgfplotsset{compat=newest}
\graphicspath{ {Images/} }
\usepackage{amsmath}
\usepackage{verbatim}
\usepackage{todonotes}
\usepackage{gensymb}
\usepackage{upgreek}
\usepackage{booktabs}
\usepackage{caption} 
\usepackage{subcaption}
\usepackage{upgreek}
\usepackage{float}
\usepackage[colorlinks=true, allcolors=blue]{hyperref}
\usepackage{threeparttable}
\usepackage{amsmath}
\usepackage{siunitx}
\usepackage{algpseudocode}
\usepackage{algorithm}
\usepackage{makecell}

\usepackage[toc]{appendix}

\title{Neural Cellular Automata for Solidification Microstructure Modelling}
\author[label1,label2]{Jian Tang}
\author[label3]{Siddhant Kumar}
\author[label2]{Laura De Lorenzis\corref{cor1}}
\author[label1]{Ehsan Hosseini \corref{cor2}}
\cortext[cor1]{Corresponding author: L. De Lorenzis,  ldelorenzis@ethz.ch, ETH}
\cortext[cor2]{Corresponding author: E. Hosseini,  ehsan.hosseini@empa.ch, Empa}
\address[label1]{Empa Swiss Federal Laboratories for Materials Science \& Technology, 8600 D\"{u}bendorf, Switzerland}
\address[label2]{Department of Mechanical and Process Engineering, ETH Z\"{u}rich, 8092 Z\"{u}rich, Switzerland}
\address[label3]{Department of Materials Science and Engineering, Delft University of Technology, 2628 CD Delft, The Netherlands}
\date{\today}
\journal{Journal Computer Methods in Applied Mechanics and Engineering}

\begin{document}

\begin{abstract}
%The computational prediction of the  microstructure of metallic systems upon solidification
%the process-microstructure-property relationship for metallic systems has gained increasing attention, particularly to enable the exploitation of 
%is of crucial importance e.g. in advanced manufacturing technologies. However, one of the current bottlenecks is the computational speed. 
%To this end, a significant enhancement of the computational efficiency of the involved simulations is required. With this motivation, the present study introduced a new method, called 
%In this paper, 
We propose Neural Cellular Automata (NCA) to simulate the microstructure development during the solidification process in metals. Based on convolutional neural networks, NCA can learn essential solidification features, such as preferred growth direction and competitive grain growth, and are up to six orders of magnitude faster than the conventional Cellular Automata (CA). Notably, NCA deliver reliable predictions also outside their training range, e.g.\ for larger domains, longer solidification duration, and different temperature fields and nucleation settings, which indicates that they
learn the physics of the solidification process. While in this study we employ data produced by CA for training, NCA can be trained based on any microstructural simulation data, e.g.\ from phase-field models.
%and therefore take the role of a universal method for efficient and reliable microstructure simulation. 
\end{abstract}
\maketitle

\section{Introduction}\label{sec:Intro}

The microstructure of metals collectively denotes their grains and metallurgical phases, their size, relative amount, shape/morphology and crystallographic orientation (texture). The microstructure strongly influences the physical and mechanical response of metallic materials, and a wide range of properties can be achieved by manipulating the microstructural features through thermomechanical processing. Therefore, correlating processing conditions and the resulting microstructure has always been an important research field. Understanding the influence of the solidification process conditions on the features of the resulting microstructure has attracted even more attention since the advent of metal additive manufacturing (MAM) processes that allow site-specific control of the solidification conditions \cite{sciAMRela, sciAMexam}. However, achieving the goal of microstructure control, e.g.\ fabricating functionally graded materials, via MAM requires reliable solidification microstructure models that enable quantitative prediction of the process-microstructure relationship.

Microstructure formation during the solidification process comprises two phenomena:\ nucleation and grain growth \cite{kou2003welding}. Nucleation is the formation of stable nuclei in the molten metal and typically requires that the melt cools down below its melting temperature (undercooled). Growth includes enlargement of the nuclei (or existing grains) depending on the temperature field and microstructure state in the vicinity of the solidification interface \cite{kou2003welding,kurz1989fundamental,kurz1986theory,lipton1984dendritic}. The formation of solid material during growth is accompanied by the release of latent heat of fusion which should be extracted from the solidification interface to allow further growth. From a thermal point of view, this indicates that the growth direction aligns with the maximum temperature gradient direction, where the fastest heat dissipation through conduction takes place \cite{kou2003welding}. While crystallographic orientation considerations are not relevant for nucleation and formed nuclei have random orientations \cite{rappaz1993probabilistic,nuc}, the growth phenomenon is anisotropic and nuclei/grains grow faster along certain crystallographic orientations (i.e.\ \emph{preferred growth directions}, $<$100$>$ for BCC/FCC crystals) \cite{kou2003welding}. Thermal and crystallographic factors collectively result in the selective growth of nuclei/grains or the so-called \emph{competitive grain growth} phenomenon, which means that the nuclei/grains whose preferred growth directions are aligned with the maximum temperature gradient grow faster and become the dominant feature of the microstructure \cite{kou2003welding}. Consideration of the above phenomena is the essential part of the solidification microstructure modelling strategies \ \cite{kou2003welding,Basak2016, MC, PFsol1, PFden1,YanPFGPU,npj_pf,gandin19973d,gandin1999three, ca_cast1}.

Besides simple analytical/empirical models \cite{kou2003welding,Basak2016, MC}, the two 'physics-based' microstructure modelling techniques are Phase Field Modeling (PFM) and Cellular Automata (CA). The main focus of these two modelling strategies is on representing the grain growth phenomena, while they often adopt empirical probabilistic or ad-hoc considerations for the nucleation phenomena \cite{rappaz1993probabilistic,kundin2015phase}.  PFM for microstructure modelling \cite{PFcon1,PFcon2} represents the microstructure by a set of conserved, e.g.\ for solute atom concentration, and non-conserved phase field parameters, e.g.\ for the state (solid/liquid) and grain ID (grains with various crystallographic orientation). PFM defines the free energy of the system as a function of the conserved and non-conserved phase field parameters, their gradients, temperature, and possibly additional variables. An anisotropic grain boundary energy is considered to account for the preferred growth direction. This energy term is maximum for grain boundaries perpendicular to the crystal preferred growth directions so that grains tend to grow anisotropically along these directions \cite{PFcon2,PFCON3}.
The Cahn-Hilliard \cite{CahnHall1,CahnHall2} and Allen-Cahn \cite{PFcon1,AllenCahn1, AllenCahn2} equations are respectively used to govern the evolution of conserved and non-conserved phase-field parameters such that the free energy of the system tends to its minimum. The kinetics of the evolution depends on the reduction rate of the system's free energy and on temperature-dependent mobility factors. PFM is a very detailed microstructural modelling strategy that not only well represents important phenomena such as preferred growth directions and competitive grain growth, but also can provide predictions about microscale features such as secondary dendrite growth morphology and solute atom distribution \cite{PFsol1, PFsol2, PFden1}.  However, the need for space and time resolutions in the orders of 0.01-1 \micro m and 0.1-10 \micro s leads to a very high computational cost for PFM, which limits the simulation domain size to hundreds of micrometres \cite{PFsol1, PFsol2, PFden1, liu_pf, liu_pf2, PFdom2, YanPFGPU,npj_pf}. As an example, Yan et al.\ \cite{YanPFGPU} report a computational time of 13 days for the solidification microstructure simulation of a 350$\times$350$\times$150  \micro m$^3$ domain using an NVIDIA Tesla M2090 GPU.  

The CA method, developed by Gandin and Rappaz \cite{gandin19973d,gandin1999three}, evaluates the microstructure and temperature state locally in the vicinity of the solidification interface and combines crystallographic considerations and experimentally or theoretically driven dendrite tip growth rate equations to predict the grain growth phenomenon \cite{kurz1989fundamental,kurz1986theory,lipton1984dendritic}. Numerous examples of the application of CA for investigating the microstructure development during casting \cite{gandin1999three,ca_cast1,ca_casting2} and MAM \cite{lian2019cellular, KOEPFca,ma1AMCA} demonstrate the experimental relevance of the CA method and its ability for representing the preferred grain growth and competitive grain growth phenomena. CA is often used to predict grain size, morphology, and crystallographic texture \cite{lian2019cellular, mohebbi2020implementation,KOEPFca}, although there are a few studies that extend CA for predicting solute atom distribution \cite{WANGCA, RolCA}. The typical space and time resolutions for CA simulations are 0.1-10 \micro m and 0.5-100 \micro s, respectively. In comparison with PFM, the coarser discretization and the rather local and simple calculations lead to lower computational cost for CA, such that the simulation domain size can be up to several cubic millimetres for a few days of computations \cite{lian2019cellular,mohebbi2020implementation, ma1AMCA,KOEPFca}. Still, it is obvious that even CA cannot be used for large-scale simulations or, importantly, for optimization problems where numerous simulation runs are required to find the optimum set of process conditions for a desired type of microstructure. Therefore, there is a substantial need to develop novel microstructure simulation techniques which retain physical relevance and reliability but reduce the computational cost by orders of magnitude. 

Deep learning has recently emerged as a central tool for scientific computing, and there are many successful examples where a deep neural network (DNN) is trained to serve as a highly efficient simulation tool \cite{DL_m1,DL_cry,npj-dnn2022}. A common DNN-based strategy for microstructure modelling is to project the microstructure onto a low-dimensional space by utilizing statistical descriptors and/or dimensionality reduction techniques, and to use DNNs to describe their evolution in this latent space \cite{DL1,DL_rep,DL-sta}. However, the employment of simple statistical descriptors, such as spatial correlation functions of phases or concentration distributions, may compromise the accuracy of predictions and impede the back transformation to microstructure images \cite{DL1,DL_rep,DL-sta,twopcor1,twopcor2}. Alternatively, advanced auto-encoders can be used to reduce the dimensionality and reconstruct from the latent space back to the original microstructure with minimal loss of information \cite{DL2,DL_auto}. 
However, modelling the development of the microstructure in a latent space precludes the direct integration of physical constraints and laws into the DNN, resulting in black-box solutions with limited applicability beyond their trained domain. 

In this paper, we propose a new DNN-based approach, denoted as neural cellular automata (NCA), to predict microstructure development during the solidification process. It is similar to the classical microstructure modelling frameworks (e.g.\ CA and PF) that ‘graphically’ describe the evolution of microstructure, and allows for the direct incorporation of physical concepts. Our NCA idea is inspired by the work of Mordvintsev et al.\ \cite{nca1,nca2}, who integrated a fully connected neural network into the CA framework to reconstruct or repair dedicated image patterns seen during the training. Similar to their work \cite{nca1,nca2}, our NCA is an extension of the conventional CA, where a convolutional neural network (CNN) acts as a highly efficient and flexible 'brain' for governing the evolution of cells.

%\section{Result}\label{sec:result}
\section{From Cellular Automata (CA) to Neural Cellular Automata (NCA)}\label{sec:architecture}
The CA microstructural modelling framework discretizes the simulation domain into a series of cubic/square cells, each containing a set of microstructure-related parameters such as phase state, crystallographic orientation, temperature, and possibly more \ \cite{KOEPFca,lian2019cellular}. The phase state can be either solid (S), liquid (L), or growing (G). The (G) cells are located between the (S) and (L) cells, representing the solidification interface. Three Euler angles describe the crystallographic orientation of (S) and (G) cells; these are angles between the global reference coordinate system and the principal directions of the crystal at the cells (i.e.\ $<$100$>$ crystallographic orientations, which for BCC/FCC coincide with their preferred growth directions). A number of simple rules then govern the evolution of the parameters of each cell, based on information from the cell itself and from those in its local neighbourhood. The rules are devised such that the physics of grain growth during solidification is reproduced. For example, a (G) or (S) cell with a temperature higher than the melting point transforms to the (L) state, or a (G) cell with no (L) cell in the neighbourhood changes to the (S) state. As essential constituent, CA employ the so-called 'decentred octahedron method' \cite{gandin19973d,gandin1999three} to consider the preferred growth directions. For BCC/FCC systems, the 'decentred octahedron' method assumes an octahedron for each (G) cell whose diagonals are parallel to the $<$100$>$ directions of the cell's crystal. A theoretically or empirically driven equation for dendrite tip growth rate takes the cell's temperature and defines the growth of the octahedron along its diagonals. The octahedron envelope finally grows into neighbouring cells, and when it captures the centre of a neighbour liquid cell, its state (L) transforms to (G). The new growing cell inherits the same crystallographic orientation as the parent cell and starts to grow its own octahedron. The growth of octahedra along $<$100$>$ crystallographic orientations and the dependency of their growth rate on the local temperature ensure that CA reproduce the phenomena of preferred growth directions and competitive grain growth in solidification microstructure modelling. The 'decentred octahedron method' schematic and the detailed algorithm of CA are given in \textbf{\ref{sec:traindata}}. 

\begin{figure}[!ht]
	\centering
	\includegraphics[width=0.65\linewidth]{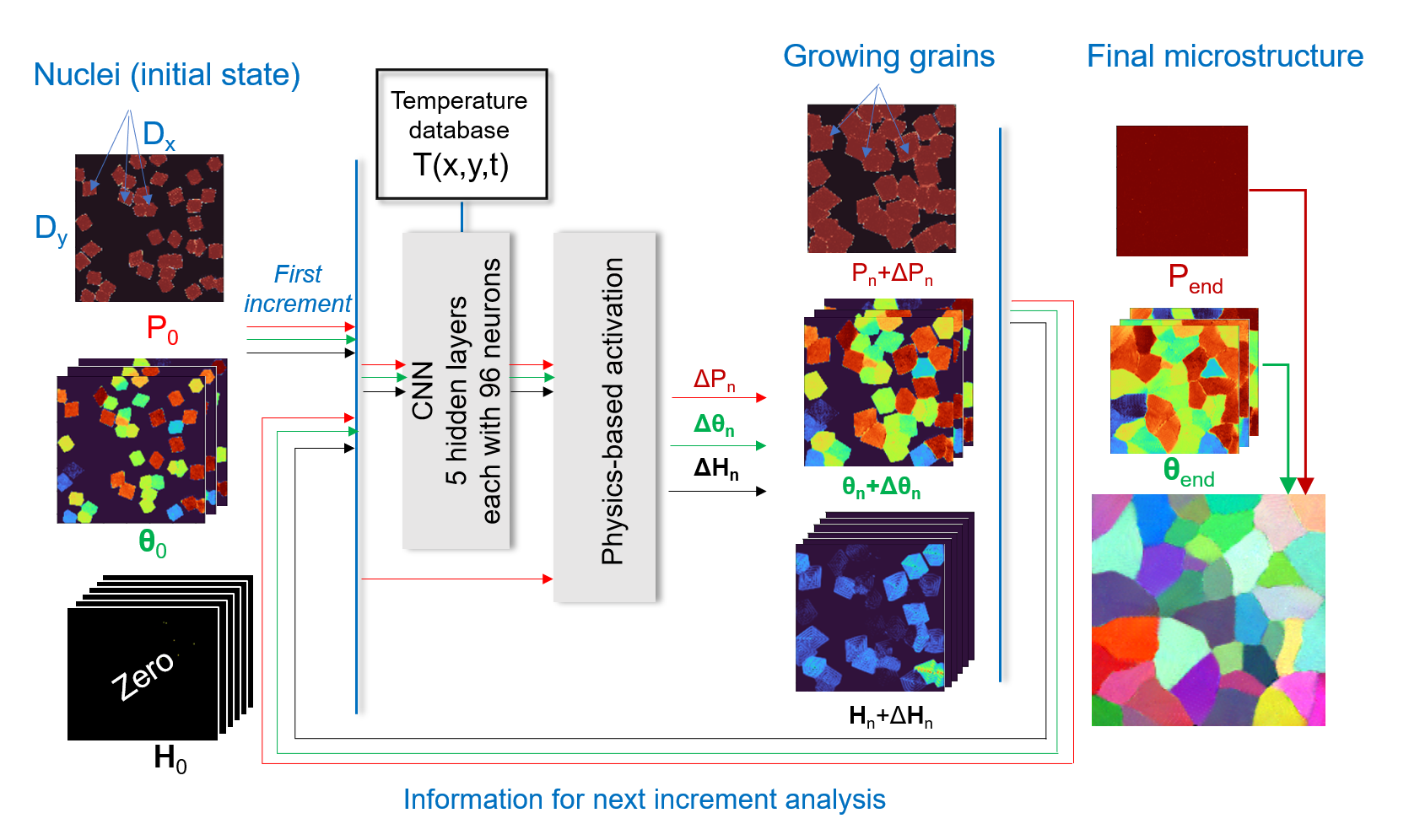}
	\caption{Schematic of the NCA modelling strategy.}
	\label{fig:Schematics}
\end{figure}

There are factors which constrain the computational efficiency of CA. The incremental progress of CA simulations should be small to ensure that the 'capture' of the neighbouring cells - i.e. the transformation of the (L) neighbouring cells into the (G) state due to growth - is accurately modelled. The time increment is often limited to $\Delta t = \alpha d/v_{max}$, where $d$: cell size, $\alpha$: a constant between $(0,1]$, and $v_{max}$: maximum growth rate among all growing cells in the simulation domain. Additionally, the consideration of information from neighbouring cells is typically limited to their immediate surroundings and is done through a 'for loop' method.
%In cellular automata (CA) programs, the consideration of information from neighboring cells is limited to their immediate surroundings and is often done by using a for loop. 
%from a programming point of view, assessing the information of the neighbouring cells is often performed only for their immediate neighbourhood and through computationally less efficient 'for loop'.
NCA enhances the computational efficiency and, potentially, the flexibility and physical relevance of CA through the power of CNNs.

\textbf{Figure~\ref{fig:Schematics}} illustrates a schematic representation of the NCA modelling strategy. Similar to CA, NCA discretise the simulation domain into cubic/square pixels and incrementally (i.e. frame after frame) update the microstructure based on the temperature and microstructure states in the local neighbourhoods of individual pixels. The required input data to initialise NCA are the locations of the nuclei, their size and their crystallographic orientation. Also required is the temperature evolution within the simulation domain during the solidification process; the temperature data are fed to the network as undercooling magnitude  (i.e.\ $\Delta T=T_{melting}-T$). A continuous variable $P \in$ [0,1] is considered for each pixel to represent its phase state, where $P = 0$ and $P = 1$ indicate fully liquid and solid states, respectively. Similar to CA, the crystallographic orientations of the pixels are defined with three Euler angles $\boldsymbol{\theta} =(\alpha_1, \alpha_2, \alpha_3)$, i.e.\  the angles between the principal directions of the solidified pixel and the reference coordinate system. 
%The output of NCA is then frame-by-frame microstructure development during the solidification process in terms of pixel-wise evolution of phase state and Euler angles. 

NCA predicts the sequential microstructure evolution by recurrently employing a trained CNN. The CNN takes the temperature, phase state, three Euler angles and six hidden channels as inputs and delivers pixel-wise increments of the phase state, Euler angles and hidden channels (i.e.\ $\Delta$P, $\Delta\theta$ and $\Delta$H, respectively) as output. Apart from the phase state and Euler angles, the hidden channels allow the CNN to exchange information between different frames and better track the evolution of microstructural features.  The physical meaning of the hidden channels is not explicitly defined and they have zero value at the start of the analysis. The number of hidden channels and the CNN architecture are decided based on a hyperparameter tuning exercise (see \textbf{Section \ref{sec:hyper}}). Accordingly, six hidden channels and 1+5+1 convolutional layers with ReLU activation function and 96 neurons per hidden layer, including 3×3 input/hidden-layer kernels and 1×1 output-layer kernels, are utilized in the NCA. The details of the adopted CNN and NCA training are given in \textbf{Section \ref{sec:cnn_Archi}} and \textbf{\ref{sec:train}}, respectively.

To improve the physical relevance of the algorithm, the NCA integrates a 'physics-based activation function' at the output layer to prevent the CNN from violating explicit known physical laws of solidification. This 'activation function' directly encodes the knowledge that the grain growth mainly takes place at the solidification interface; it enforces no change of microstructure states for a pixel when: i) $P <$ 0.1 and $T > T_{melting}$, i.e.\ the liquid pixel is at a temperature above the melting point, ii) no neighbour pixel with $P \geq$ 0.1, i.e.\ not in the vicinity of the solidification interface, iii) $P > 0.99$ and $T < T_{melting}$, i.e.\ the solid pixel temperature is below the melting point. Values of 0.1 and 0.99 are the thresholds for the phase state $P$ indicating liquid and solid. They can be seen as network hyperparameters and are taken from \cite{nca1,nca2}.
 
%\textcolor{red}{As the grain growth mainly happens at the solidification interface, a filter is integrated into the NCA model to add known physics into the trainable CNN. }The filter enforces no update of a cell when: i) $P <$ 0.1 and $T > T_{melting}$, i.e.\ the liquid pixel is at a temperature above the melting point, ii) no neighbour pixel with $P \geq$ 0.1, i.e.\ not in the vicinity of the solidification interface, iii) $P > 0.99$ and $T < T_{melting}$, i.e.\ the solid pixel temperature is below the melting point. \textcolor{red}{This filter can be considered as an "activation function" of the CNN output, which is defined based on the pixel phase state P of CNN inputs. 0.1 and 0.99 are the criteria phase states for liquid and solid, which are hyperparameters chosen based on literature values \cite{nca1,nca2}.} %Finally, the microstructure frames can be obtained by assembling the phase state and three Euler angles, as shown in Fig.\ \ref{fig:Schematics}.

The adoption of multi-layer convolution enables NCA to implicitly consider information from large neighbourhoods and integrate complex and highly nonlinear rules for governing the evolution of the microstructural parameters of the pixels. This leads to an advantage of NCA over CA in encoding the complexity of the solidification process, e.g. when an NCA model is trained based on data from PFM. \textbf{Section \ref{sec:computation}} also discusses the significantly higher computational speed of NCA over CA.    

\section{Methods}\label{sec:Methods}
\subsection{Materials}\label{sec:CA}
In this work, we choose Hastelloy X as a model alloy to investigate the applicability of NCA in predicting its microstructure development under various solidification conditions. The generation of NCA training and validation data through CA necessitates information about the preferred growth direction and dendrite growth kinetics. 
 Hastelloy X is a nickel-based alloy with an FCC crystal structure (solid-solution strengthened) and hence has the preferred growth direction along the $<$001$>$ orientation \cite{AM_HX2,AM_HX1,kou2003welding}. Meanwhile, the Kurz-Giovanola-Trivedi (KGT) model \cite{kurz1986theory} is used to define the dendrite tip growth rate of Hastelloy X as a function of undercooling as: 
\begin{equation}\label{eq:1}
v(\Delta T) = 4.60 \times 10^{-10} \times \Delta T^2 + 3.15 \times 10^{-6} \times \Delta T^3
\end{equation}
Further details about the derivation of the above KGT model for Hastelloy X are described in \textbf{\ref{sec:dgv}}.

\mbox{}

\subsection{Convolutional neural networks for NCA}\label{sec:cnn_Archi}
\begin{figure}[!th]
	\centering
	\renewcommand{\thefigure}{\arabic{figure}}
	\includegraphics[width=0.85\linewidth]{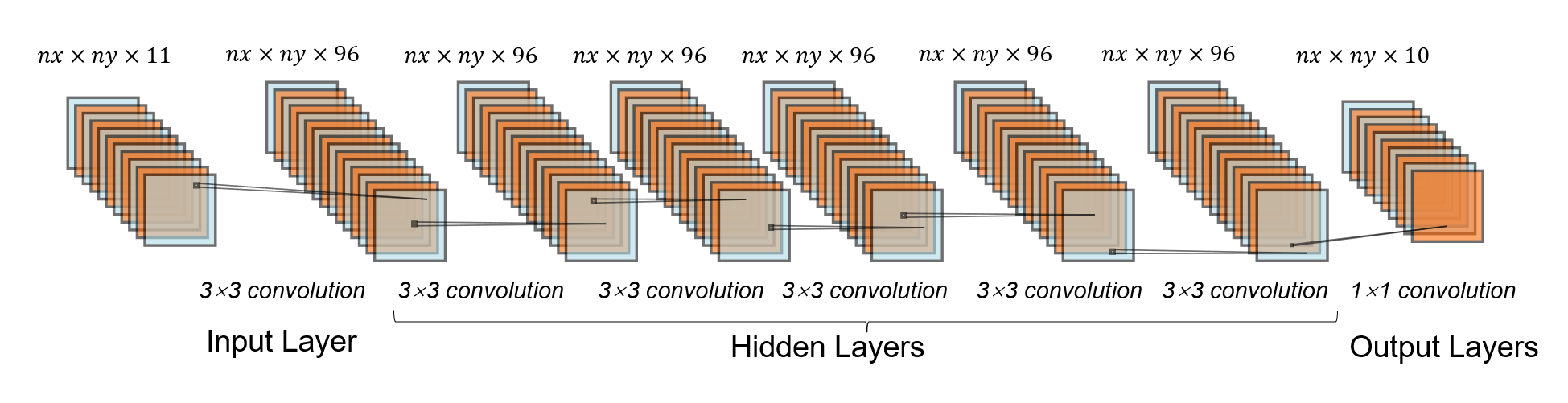}
	\caption{Schematic of selected CNN architecture for NCA (image generated by the online tool \href{http://alexlenail.me/NN-SVG/AlexNet.htm}{NNSVG}). The numbers at the top represent the dimensions of the input/output for each convolutional layer, and the bottom numbers indicate the size of each convolutional kernel.}
	\label{fig:cnn_archi}
\end{figure}

The architecture of the CNN in NCA is composed of fully connected convolutional layers, as depicted in \textbf{Figure \ref{fig:cnn_archi}}. Each convolutional layer consists of a kernel that performs pixel-wise convolution. The mathematical transformation in a convolutional layer is:
\begin{equation}\label{eq:2}
\textbf{x}_{i\!jk}^{l+1} = f(\sum_{o}\sum_{p}\sum_{q}{\textbf{w}_{o\!pqk}^{l} \textbf{x}_{i\pm o,j\pm p,q}^{l}} + \textbf{b}_k^l)
\end{equation} % we need to discuss the formula
where $\textbf{w}$ and $\textbf{b}$ denote the trainable convolutional kernel weights and bias, $f$ represents a non-linear activation function, $\textbf{x}^l$ is the input of the current convolutional layer and $\textbf{x}^{l+1}$ is its output, which serves as the input for the next layer. 
%$\textbf{x}_l^{nx\times ny\times inp}$ represents the input, $\textbf{w}_l^{ks\times ks\times inp \times out}$ is the convolution kernel weights, $\textbf{b}_l^{out}$ is the bias, and $\star$ denotes the convolution operator. In this equation, $ks$ is the kernel size, while $inp$ and $out$ are the numbers of channels for input and output matrices of the convolutional layers, respectively. The output passes through a non-linear activation function $f$ before being used as inputs for the next layer.

In brief, CNN can be considered a neural network 'function' that decides the evolution in the state of a pixel by analysing the information from its neighbouring pixels.
The weights and biases of the network are determined through network training, where a high-dimensional optimization problem is solved to minimize the difference between the CNN predictions and the ground truth data generated by the CA (as explained in \textbf{Section \ref{sec:train}}). 
To achieve the best performance, the hyperparameters of CNN, such as the number of hidden layers and the activation function type, need to be optimized (as discussed in \textbf{Section \ref{sec:hyper}}).

\subsection{NCA training \& testing}\label{sec:train}
Table \ref{tab:traindata} shows that different variants of NCA are trained to represent Hastelloy X solidification microstructure under different conditions. Ground truth data are created using CA simulations with a domain size of 55×55 $\micro$m$^2$ for each solidification condition. The data are divided into a 9:1 ratio for training and validation  \cite{datasplit1,datasplit2}. Additionally, ten further CA simulations are performed to evaluate the generalization error of NCA variants under unseen cases, e.g.\ larger domain sizes, longer solidification durations, different temperature fields and nucleation settings. The orientation and location of nuclei are randomly assigned in the CA simulations for generating training, validation, and testing data except explicitly mentioned otherwise (i.e.\ single grain growth). The CA simulations adopt a time increment of $\Delta t = 8\ \micro$s, except for quasi-3D cases, where $\Delta t = 40\ \micro$s.  %The nuclei Euler angles were selected from 18 and 729 unique crystallographic orientations for 2D and quasi-3D simulations, respectively.

During training, the CA-predicted phase state and Euler angles, along with undercooling values, are used as inputs for the NCA (the Euler angles and undercooling are normalized by 90° and 100 K, respectively). The CNN is initialized with zero biases and random weights and kernel parameters (except for the output layer).  The training uses the Adam optimizer (learning rate of $10^{-3}$ decaying to $9\times 10^{-5}$) for 7000 epochs and the L2 loss defined by equation \ref{eq:3}. To achieve faster convergence and following \cite{nca1,nca2}, the weight and bias of the output layer are initialized with zero and the gradients of the loss are normalized by their Frobenius norms at every epoch (to prevent gradient explosion).

\begin{equation}\label{eq:3}
%\begin{split}
l=\sum_{F\!rame} \left[ \sum_{S\!olid pixel} \left[(P^{NCA}\!-\!P^{CA})^2\!+\|\boldsymbol{\theta}^{NCA}\!-\!\boldsymbol{\theta}^{CA}\|^2 \right]/N_{S\!olidpixel} \right] / N_{F\!rame}
%\end{split}
\end{equation}
Here $N_{F\!rame}$ and $N_{S\!olidpixel}$ are the numbers of the time frame and solid pixel in each frame (pixel with $P>0.99$), respectively. Finally, NCA's accuracy is evaluated by comparing its results with CA predictions:
\begin{equation} \label{eq:4}
Accuracy= 1- N_{\Delta\phi(CA,NCA)>15\degree}\//N_{total}  
\end{equation}
where $N_{total}$ is the total number of pixels and $N_{\Delta\phi(CA,NCA)>15\degree}$ is the number of pixels with orientation error $\Delta\phi$ exceeding 15° (i.e.\ the threshold of establishing grain boundary). 

\begin{table}[h!]
\centering
\scriptsize
{%
\begin{threeparttable}
\caption{Details of the NCA training for different solidification conditions.}\label{tab:traindata}
\begin{tabular}{llll}
\toprule
NCA model variant & Single-grain growth\tnote{1} & Multi-grain growth & Training time\tnote{2}\\
\midrule
Isothermal 2D single-grain growth\tnote{3} & 18 CA simulations & - & 0.8 h\\
Isothermal 2D multi-grain growth & 18 CA simulations	& 27 CA simulations	& 2 h\\
Isothermal quasi-3D multi-grain growth & 729 CA simulations &	929 CA simulations	& 72 h\\
Non-isothermal 2D multi-grain growth \tnote{4} & 180 CA simulations & 270 CA simulations & 24 h\\
\bottomrule
\end{tabular}
\begin{tablenotes}
        \scriptsize
        \item[1] The domain size is 55$\times$55 $\micro m^2$ and the nucleation density for the multi-grain simulations is 8260-11570 mm$^{-2}$. The nuclei locate at the centre of the domain for the single-grain growth samples. It is observed that including single-grain growth data enhances the training efficiency for representing multi-grain growth (See \textbf{\ref{sec:traindata}}). 
        
        \item[2] An NVIDIA RTX A5000 GPU is employed. 
        
        \item[3] A constant undercooling of 20 K is used for isothermal simulations.  

        \item[4] {Non-isothermal simulations include solidification with 45 different nuclei settings (including single and multi-nuclei) under 10 different temperature fields with undercoolings in the range of 20-25 K.}

\end{tablenotes}
\end{threeparttable}}% 
\end{table}

\subsection{Hyperparameter tuning}\label{sec:hyper}
A hyperparameter tuning exercise is conducted to find the optimal CNN architecture. This involves training and evaluating multiple neural networks with different architectures, varying the number of hidden convolutional layers and neurons per hidden layer. We keep other details, such as kernel size, learning rate, and activation function, consistent with references \cite{nca1,nca2}.

The hyperparameter tuning exercise is conducted solely for non-isothermal solidification data, and the CNN architecture that yields the best results is used for all other conditions. To reduce the computational costs of hyperparameter tuning, we limit the number of epochs for training each model to 2500 and use a fast-decaying learning rate schedule (70\% decay every 1000 epochs). This approach ensures that each model is trained in less than 0.5 hours.

Table \ref{tab:hyper} shows that the best training and validation accuracy is achieved with five hidden layers and 96 neurons per hidden layer and hence is adopted for this study. The impact of varying the number of input hidden channels on CNN accuracy is also examined which indicates that the adoption of six hidden channels produces the best results (Table \ref{tab:hidchan}).

\begin{table}[h!]
\centering
\scriptsize
\renewcommand{\thetable}{\arabic{table}}
\caption{Training and validation accuracies of the NCA model with various numbers of hidden layers and neurons per hidden layer.}\label{tab:hyper}
\setlength\tabcolsep{3pt}
\begin{tabular}{cccc}
\toprule
%\makecell[c]{Number of\\hidden layers}&\makecell[c]{Number of\\neurons per hidden layer}&\makecell[c]{Training\\accuracy (\%)}&\makecell[c]{Validation\\accuracy (\%)}\\
Number of hidden layers&Number of neurons per hidden layer&Training accuracy (\%)&Validation accuracy (\%)\\
\midrule
9 &	32 & 95.0 &	93.5\\
12 &32 & 95.0 &	93.3\\
9 &	64 & 96.1 &	93.2\\
12 &64 & 95.7 &	92.9\\
3 &	96 & 96.8 &	95.1\\
5 &	96 & 96.7 &	95.7\\
7 &	96 & 96.0 &	94.4\\
\bottomrule
\end{tabular}
\end{table}

\begin{table}[h!]
\centering
\scriptsize
\renewcommand{\thetable}{\arabic{table}}
\caption{Training and validation accuracies of the NCA model with various numbers of hidden channels.}\label{tab:hidchan}
\begin{tabular}{ccc}
\toprule
Number of hidden channels &	Training accuracy (\%) &	Validation accuracy (\%)\\
\midrule
0& 96.6& 96.1\\
3& 97.3& 96.1\\
6& 97.1& 96.4\\
9& 97.1& 96.3\\
\bottomrule
\end{tabular}
\end{table}

\section{NCA results}\label{sec:accuracy}
%This section evaluates the accuracy of the proposed NCA in predicting the microstructure formation of an FCC alloy.  For different solidification conditions, the NCA model is trained and validated using data from a series of CA simulations and then tested for unseen cases, e.g.\ for larger domain sizes or different temperature fields. 
%The corresponding accuracies are measured to evaluate the performance of trained NCA. 
%More details about the chosen model alloy (Hastelloy X) and the procedure for NCA training, validation and testing are given in \textbf{Methods}.

\subsection{Isothermal solidification: 2D single-grain growth }\label{sec:2Dsingle}
\begin{figure}[!t]
	\centering
	\includegraphics[width=0.45\linewidth]{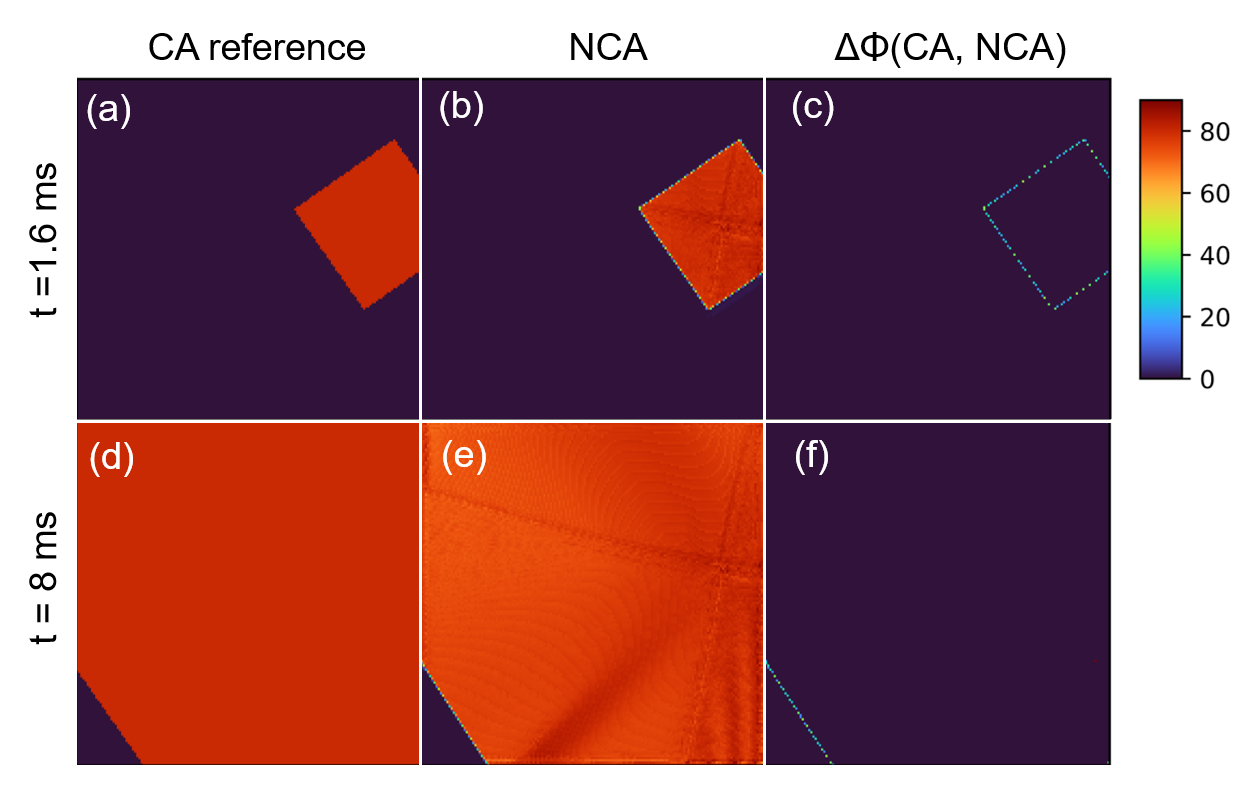}
	\caption{Test case for 2D isothermal single-grain growth in a domain of 160$\times160 \; \micro$m$^2$ with a randomly selected nucleus location: CA results (a and d), NCA results (b and e), and orientation difference $\Delta\phi$ between them (c and f). Note that the domain size and time duration of the test are 3$\times$ larger and 5$\times$ longer than those adopted for the training.}
	\label{fig:2Dsingle}
\end{figure}

Initially, the NCA model is trained with 2D CA isothermal simulation data for a domain size of 55$\times$55 $ \micro$m$^2$ (55$\times$55 pixels). A single nucleus with a random $\alpha_1$ and $\alpha_2$ = $\alpha_3$ = 0 is placed in the centre of the domain with a uniform and steady temperature field representing an undercooling of 20 K. The model achieves training and validation accuracies of 98.0\% and 97.7\%, respectively, which indicates that it learns well the kinetics of grain growth and the preferred growth direction mechanism. Furthermore, the trained model is used to predict the solidification microstructure for a 3$\times$ larger domain and a 5$\times$ longer solidification duration with the nucleus located off the centre. As presented in \textbf{Figure~\ref{fig:2Dsingle}}, the NCA can predict such cases with an accuracy of 99.6±0.2\%.

\subsection{Isothermal solidification: 2D multi-grain growth  }\label{sec:2Dmulti}
In the second stage, the data from additional multi-grain CA simulations are added to the previous ones for training and validation of the NCA. In these data nuclei are placed randomly in the 55$\times$55 $ \micro$m$^2$ domain with a nuclei density in the range of 8,260–11,570 mm$^{-2}$. Similar to the previous simulations, the 2D nature of the simulations allows considering $\alpha_2$ = $\alpha_3$ = 0 while $\alpha_1$ takes a random value. The assessment of the NCA model for multi-grain simulations indicates training and validation accuracies of 99.9\% and 96.5\%. The trained model further shows a test accuracy of 98.0±0.9\% in predicting multi-grain growth in a larger domain. \textbf{Figure~\ref{fig:2Dmulti}} illustrates NCA and CA results for a test case, showing that NCA well reproduces the microstructure evolution and the effect of nuclei density on the final grain size. For both single and multi-grain growth cases, the main inconsistency between NCA and CA is observed at the grain boundaries and solid-liquid interfaces. While CA represents the grain boundaries as a sharp interface, the involvement of the convolution operator in the NCA leads to a diffusive interface at the grain boundaries. 

To obtain grain size distributions from the microstructures generated by NCA (shown in \textbf{Figure \ref{fig:2Dmulti}g}), it was essential to establish a clear boundary between grain domains and hence, a post-processing algorithm was devised to address this requirement. The algorithm corrects the Euler angle of each pixel by replacing it with the Euler angle of a nearby nucleus that displays the most similar orientation. This correction serves to convert the diffuse interface into a well-defined sharp boundary. For more comprehensive information about this post-processing technique, refer to \ref{sec:postpro}.

\textbf{\begin{figure}[h]
	\centering
	\includegraphics[width=0.75\linewidth]{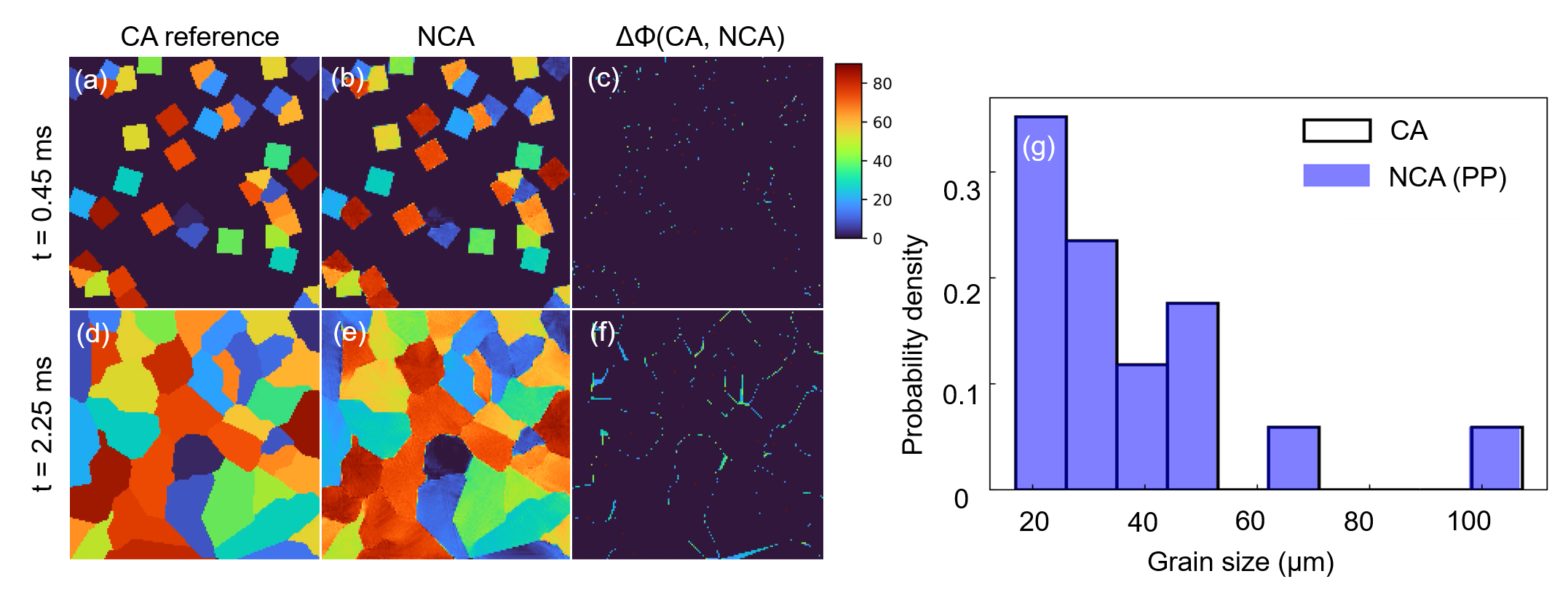}
	\caption{Test case for 2D isothermal multi-grain growth in a  domain of 160$\times160 \; \micro$m$^2$ with nucleation density of 1,562 mm$^{-2}$: CA results (a and d), NCA results (b and e), and orientation difference $\Delta\phi$ between them (c and f), grain size distributions derived from CA and NCA results (analyzed by the MATLAB toolbox MTEX \cite{mtex}) (g).}
	\label{fig:2Dmulti}
\end{figure}}

%\textbf{\begin{figure}[!ht]
%	\centering
%	\includegraphics[width=0.9\linewidth]{Figures/Multi-grain2.png}
%	\caption{Test case for 2D (a-f) and quasi-3D (g-l) isothermal multi-grain growth in a domain of 160$\times160 \; \micro$m$^2$ with nucleation density of 1,562 mm$^{-2}$: CA results (a, d, g and j), NCA results (b, e, h and k), and orientation difference $\Delta\phi$ between them (c, f, i and l). For quasi-3D cases, the three Euler angles of the nuclei are randomly selected from [0\degree, 90\degree). The RGB values of the pixel colour in (g), (h), (j), and (k) are the Euler angles normalized by 90\degree.}
%	\label{fig:multi}
%\end{figure}}

\subsection{Isothermal solidification: towards quasi-3D multi-grain growth}\label{sec:3Dmulti}
Without changing the NCA architecture, we now examine its applicability for a scenario similar to quasi-3D grain growth simulation. A quasi-3D simulation, where the 3D domain is replaced by three perpendicular 2D cross-sections \cite{quasi3d1,quasi3d2}, serves as a computationally cheaper alternative for a 3D microstructure simulation.  Such simulations require coupling between the simulations in the three perpendicular 2D cross-sections and allow for the evolution of all three Euler angles. Here, as a preliminary study, we allow for the evolution of all three Euler angles in the 2D simulations.

A new set of CA simulations including 1,658 2D runs with three evolving Euler angles for isothermal solidification with undercooling of 20 K are conducted to generate the training and validation datasets for NCA. Comparison of CA and NCA for such simulations indicates training, validation and test accuracies of 93.7\%, 87.9\% and 89.7±1.8\%, respectively. \textbf{Figure~\ref{fig:3Dmulti}} presents CA and NCA outcomes of an examined test case, showing that the trained model is capable of simulating 2D multi-grain growth with three evolving Euler angles. Compared with the previous cases in \textbf{Section \ref{sec:2Dmulti}}, a higher level of inconsistency between CA and NCA is observed, particularly at the grain boundaries. This indicates that a more complex NCA architecture may ultimately be required for quasi-3D microstructure simulations with NCA; however, this development is outside the scope of the present study. 
\textbf{\begin{figure}[!ht]
	\centering
	\includegraphics[width=0.75\linewidth]{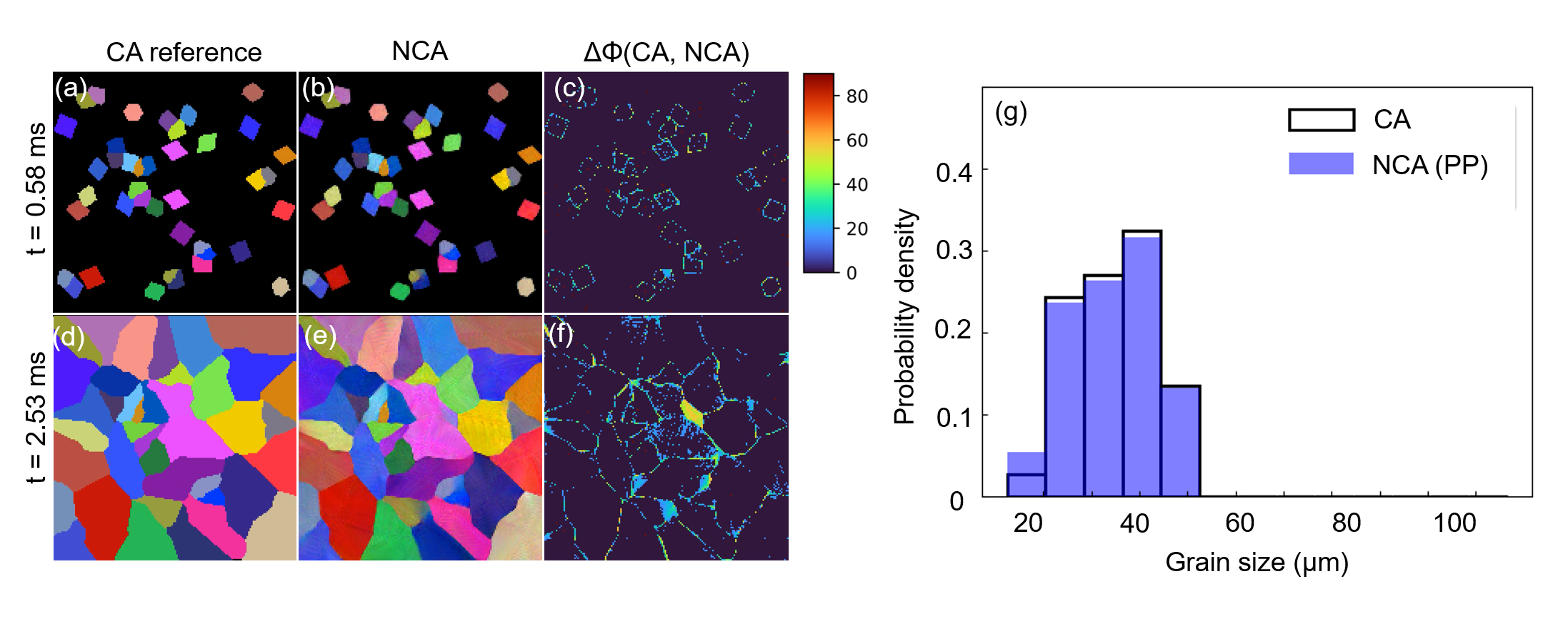}
	\caption{Test case for conditions with three evolving Euler angles in a domain of 160$\times160 \; \micro$m$^2$ with 40 nuclei: CA results (a and d), NCA results (b and e), and orientation difference $\Delta\phi$ between them (c and f), grain size distributions derived from CA and NCA results (analyzed by the MATLAB toolbox MTEX \cite{mtex}) (g). The three Euler angles of the nuclei are randomly selected from [0\degree, 90\degree). The RGB values of the pixel colour in (a), (b), (d), and (e) are the Euler angles normalized by 90\degree.}
	\label{fig:3Dmulti}
\end{figure}}

\subsection{Non-isothermal solidification: 2D multi-grain growth}\label{sec:noniso}

\begin{figure}[H]
	\centering
	\renewcommand{\thefigure}{\arabic{figure}}
	\includegraphics[width=0.75\linewidth]{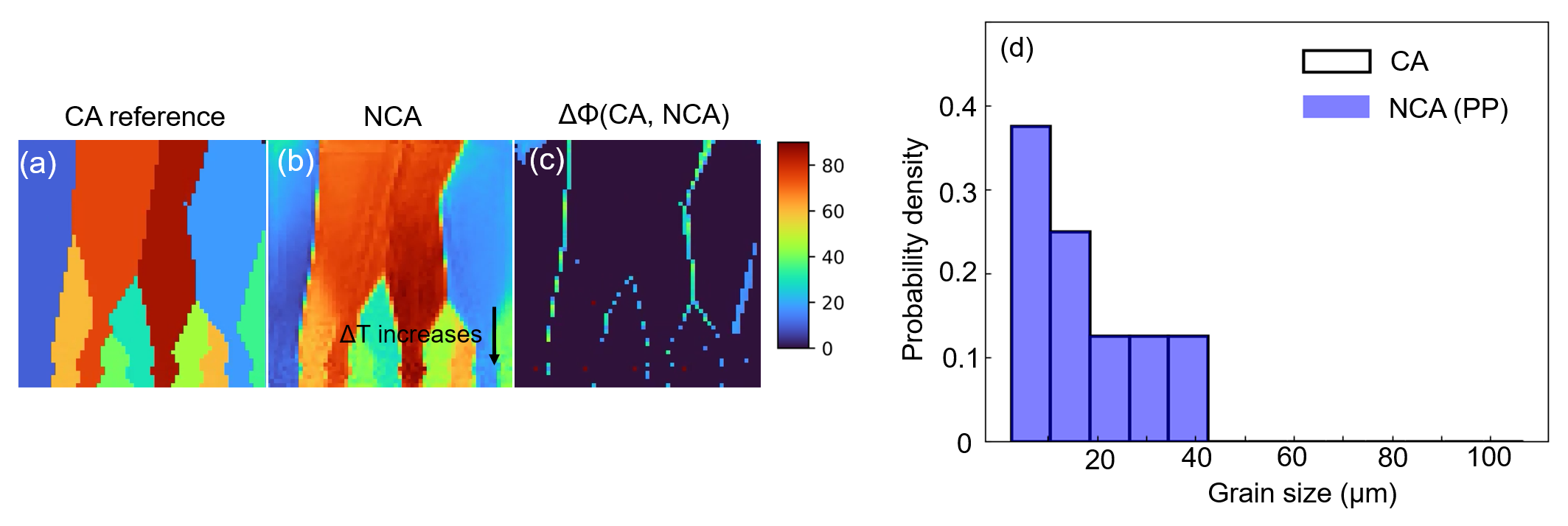}
	\caption{Test case for 2D multi-grain growth under a vertical temperature gradient with maximum and minimum undercooling of 30 and 20 K (60$\times60 \; \micro$m$^2$ domain with nucleation density of 2778 mm$^{-2}$): CA results (a), NCA results (b), and orientation difference $\Delta\phi$ between them (c), grain size distributions derived from CA and NCA results (analyzed by the MATLAB toolbox MTEX \cite{mtex}) (d). Nuclei are set 5 $\micro$m above the bottom boundary of the domain.}
	\label{fig:compgrow}
\end{figure}

For adopting the proposed simulation strategy for practical applications, we need to demonstrate its accuracy in predicting the solidification microstructure under non-isothermal conditions. In this section, the NCA model is trained based on CA simulations for 2D steady non-isothermal solidification. Accordingly, training, validation and test accuracies of 95.8\%, 95.4\% and 96.2±1.1\% are obtained, respectively. \textbf{Figure~\ref{fig:tempgrad}} shows that the NCA model predicts slower and faster growth rates for the areas with lower and higher levels of undercooling, respectively. Furthermore and as shown in \textbf{Figure \ref{fig:compgrow}}, NCA can also well predict the competitive growth mechanism, i.e.\ the grains which have one of their preferred growth directions aligned with the maximum temperature gradient grow faster than the others. 

\textbf{\begin{figure}[H]
	\centering
	\includegraphics[width=0.75\linewidth]{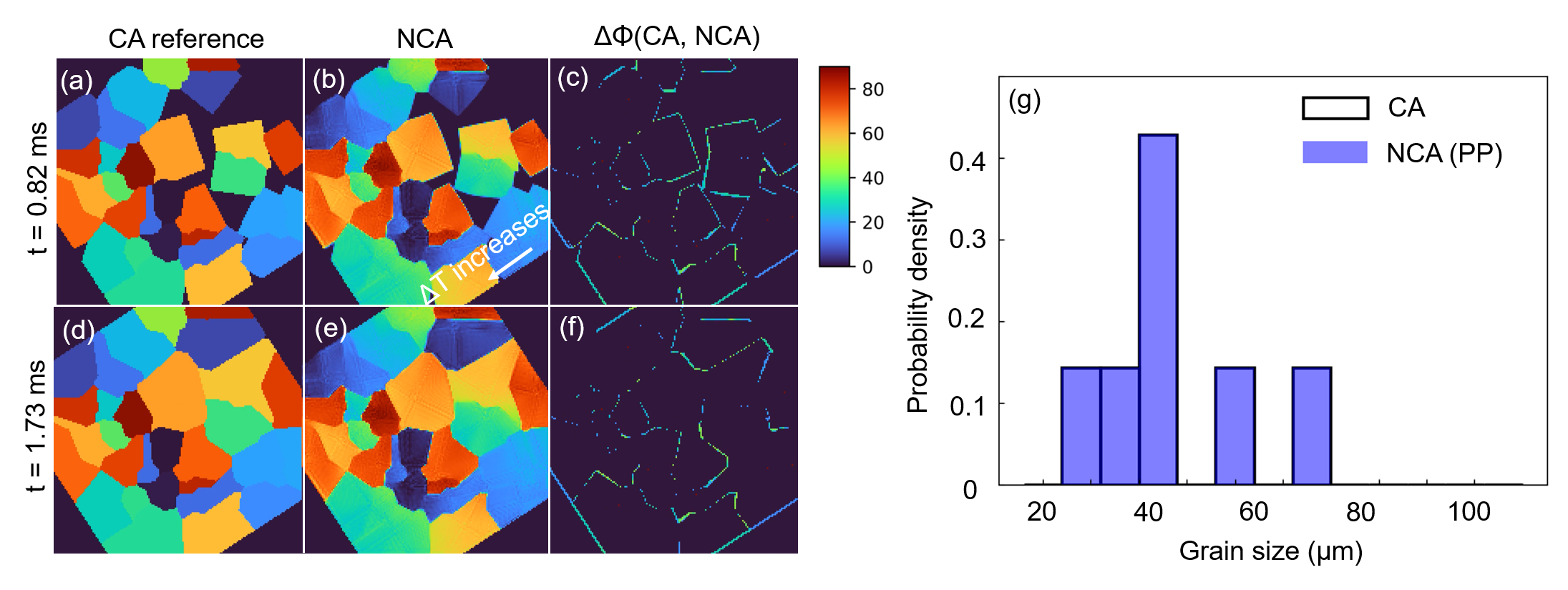}
	\caption{Test case for 2D multi-grain growth with a random temperature gradient in a domain of 160$\times160 \; \micro$m$^2$ with nucleation density of 1,562 mm$^{-2}$: CA results (a and d), NCA results (b and e), and orientation difference $\Delta\phi$ between them (c and f), grain size distributions derived from CA and NCA results (analyzed by the MATLAB toolbox MTEX \cite{mtex}) (g). A temperature gradient with maximum and minimum undercooling of 25 and 20 K is applied as shown by the arrow in (b).}
	\label{fig:tempgrad}
\end{figure}}

Without further training, the NCA model is finally used for predicting solidification under a transient non-isothermal temperature field, i.e.\ under continuous cooling. As shown in \textbf{Figure~\ref{fig:cooldown}}, although no cases with continuous cooling are included in the training, the model accurately predicts the growth kinetics and the final microstructure (test accuracy of 96.9±0.8\%).

%\textbf{\begin{figure}[!ht]
%	\centering
%	\includegraphics[width=0.9\linewidth]{Figures/Non-iso.png}
%	\caption{Test case for 2D multi-grain growth under a steady temperature gradient (a-f) and transient continuous cooling (g-l) in a domain of 160$\times160 \; \micro$m$^2$ with nucleation density of 1,562 mm$^{-2}$: CA results (a, d, g and j), NCA results (b, e, h and k), and orientation difference $\Delta\phi$ between them (c, f, i and l). A temperature gradient with maximum and minimum undercooling of 25 and 20 K is applied in (a-f) as shown by the arrow in (b). An initial undercooling of 20 K and cooling rate of $3.125\times10^3 \;$K/s is exploited for continuous cooling cases. Note that the NCA did not see any data for continuous cooling during training.}
%	\label{fig:non-iso}
%\end{figure}}

\textbf{\begin{figure}[!ht]
	\centering
	\includegraphics[width=0.75\linewidth]{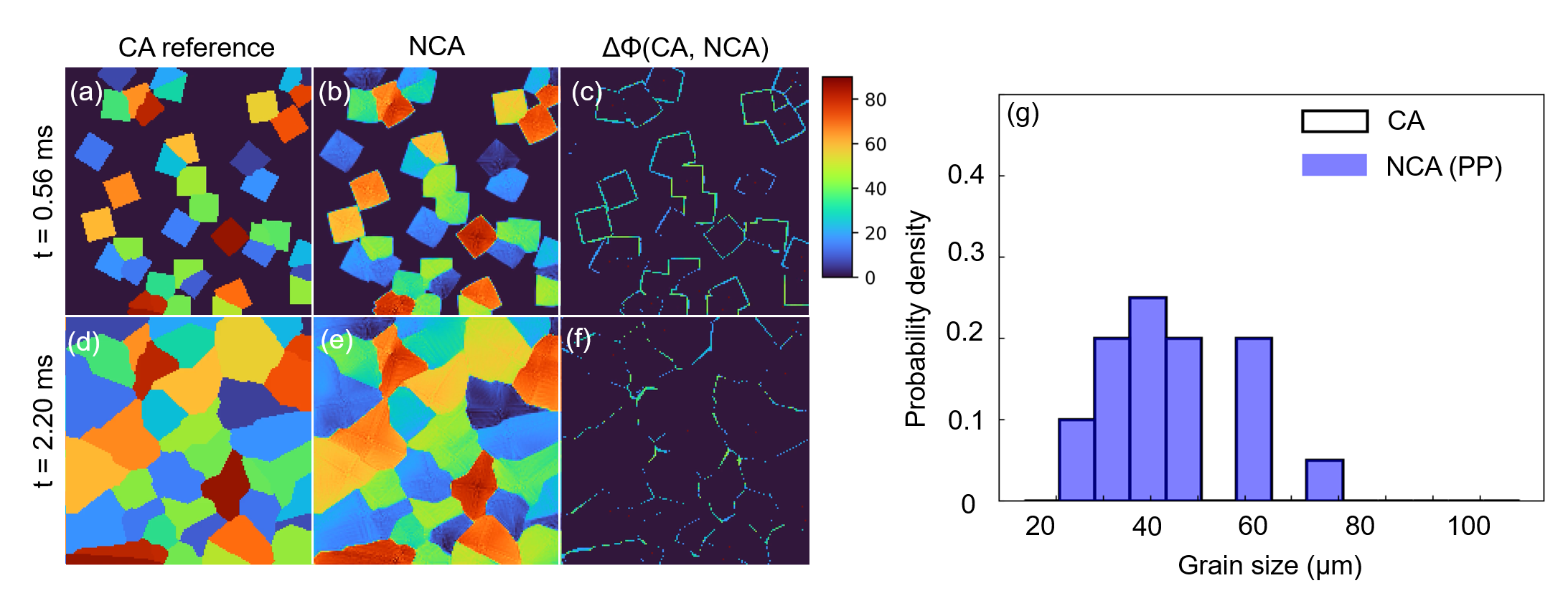}
	\caption{Test case for 2D multi-grain growth under continuous cooling with an initial undercooling of 20 K and cooling rate of $3.125\times10^3 \;$K/s in a domain of 160$\times160 \; \micro$m$^2$ with nucleation density of 1,562 mm$^{-2}$: CA  results (a and d), NCA results (b and e), and orientation difference $\Delta\phi$ between them (c and f), grain size distributions derived from CA and NCA results (analyzed by the MATLAB toolbox MTEX \cite{mtex}) (g). Note that the NCA did not see any data for continuous cooling during training.}
	\label{fig:cooldown}
\end{figure}}

In summary, the results presented in this section demonstrate the accuracy of the NCA microstructure modelling strategy and its ability to learn critical physical mechanisms of solidification such as preferred growth direction and competitive growth under various solidification conditions.  

\section{NCA computational speed}\label{sec:computation}
This section compares the computational speed of NCA and CA for a set of benchmark cases. As noted in the introduction, of the two most common physics-based microstructural modelling approaches, CA has a lower computational cost than PFM. Therefore, discussing the computational speed of NCA in comparison with CA serves as a conservative estimate for the speed increase achieved by NCA over the conventional microstructure simulation strategies. The trained NCA model in \textbf{Section \ref{sec:2Dmulti}} is used here, and its computational speed is evaluated for simulating eight different domain sizes between 40$^2$ and 880$^2$ $\micro$m$^2$ with a nuclei density of 625 mm$^{-2}$. A constant time increment of 8 $\micro$s is adopted for both CA and NCA for the primary assessments. The measured CA and NCA runtimes on Intel(R) Xeon(R) Gold 6248R CPU for various simulation domain sizes are shown as the red and yellow curves in \textbf{Figure~\ref{fig:comptime}}. An acceleration factor of 6 to 37 is observed for NCA, depending on the domain size. 

While most CA codes for microstructure modelling are only compatible with CPUs \cite{lian2019cellular,gandin19973d,KOEPFca}, NCA can easily exploit the computational power of GPUs (without extra coding).  As seen in \textbf{Figure~\ref{fig:comptime}}, GPU computation further accelerates NCA and makes them four orders of magnitude faster than CA. Notably, the runtime of NCA on GPU is not sensitive to the simulation domain size, at least within the examined range. Therefore, even higher acceleration factors can be expected for larger domain sizes.  

%\textbf{\begin{figure*}[t]
%	\centering
%\includegraphics[width=0.7\linewidth]{Figures/Comp_acc_time.png}
%	\caption{(a) Runtime of CA and NCA models for simulating microstructures with eight different domain sizes (initial nucleation density of 625 mm$^{-2}$). The time for I/O processing is not included in the runtime. The maximum error for all the presented NCA simulations remains below 8\%. (b) Accuracy of CA and NCA results with different acceleration factors of K=1, 20, 50; microstructure of 2D multi-grain growth with K=20 from CA (c) and NCA (e); orientation difference $\Delta\phi$ of CA  (d) and NCA results (f) with K=20 with respect to CA reference (K=1).}
%	\label{fig:com-acc-time}
%\end{figure*}}

Adopting a larger time increment of the simulation reduces the computational cost for both NCA and CA. To evaluate how much this deteriorates accuracy, NCA is trained for a time increment increased by a factor of K in the range of 1-50, based on a subset (one frame every K) of the CA simulation database described in \textbf{Section \ref{sec:2Dmulti}}. Furthermore, the time increment for CA is similarly increased K times and the corresponding results, along with the results of NCA, are compared with the reference CA results (those for K=1). \textbf{Figure~\ref{fig:incre}} shows the accuracy of both CA and NCA for different acceleration factors in comparison with the results of the reference CA simulation (with K=1). Expectedly, the accuracy of CA significantly decreases with the increase of K. Notably, NCA are found to be very accurate up to the acceleration factor of 20 since the multi-layer CNN implicitly considers a larger neighbourhood and utilises more complex functions for grain growth prediction than CA.

The accuracy of the NCA model drops for larger K values (e.g.\ K=50). However, as shown in \textbf{Figure~\ref{fig:incre}}, employing a more complex CNN architecture (e.g.\ with a 9×9 convolution kernel for the input layer or with 9 hidden convolution layers) leads to an excellent accuracy even for K=50. The ability of NCA to accurately reproduce the microstructure evolution using significantly larger time increments than CA originates from adopting a multi-layer convolution operator in NCA, which delivers the change in the pixel states based on information from a larger neighbourhood. These results prove that NCA is able to release the requirement for small time increments which limits the computational speed of conventional microstructure modelling techniques; hence, NCA can significantly outperform the computational speed of such simulations. 

To summarize, NCA exhibits up to six-orders-of-magnitude improvement in computational efficiency compared to CA (see \textbf{Figure \ref{fig:comptime}}). The fast computational speed of NCA comes from three contributions. Firstly, NCA replaces the 'for' loop in conventional CA with efficient convolutions, leading to over 10 times speed-up without GPUs. Secondly, the high efficiency of GPUs for parallel computation further accelerates the NCA by around three orders of magnitude. Finally, NCA can benefit from the possibility of adopting larger time increments, up to 50 times larger than the conventional CA in the shown cases.
\textbf{\begin{figure}[H]
	\centering
	\includegraphics[width=0.45\linewidth]{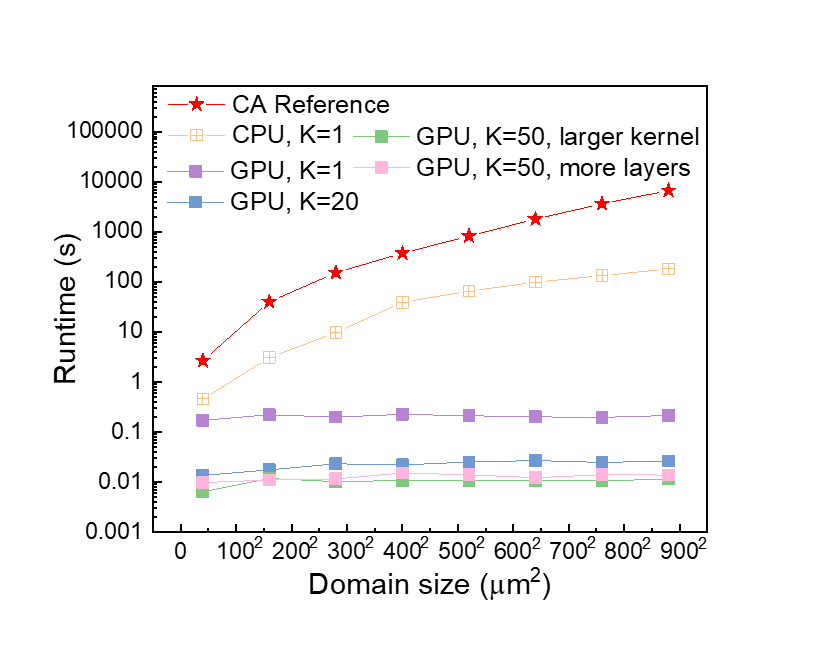}
	\caption{Runtime of CA and NCA models for simulating microstructures with eight different domain sizes (initial nucleation density of 625 mm$^{-2}$). The time for I/O processing is not included in the runtime. The maximum error for all the presented NCA simulations remains below 8\%.}
	\label{fig:comptime}
\end{figure}}
\textbf{\begin{figure}[H]
	\centering
        \includegraphics[width=0.6\linewidth]{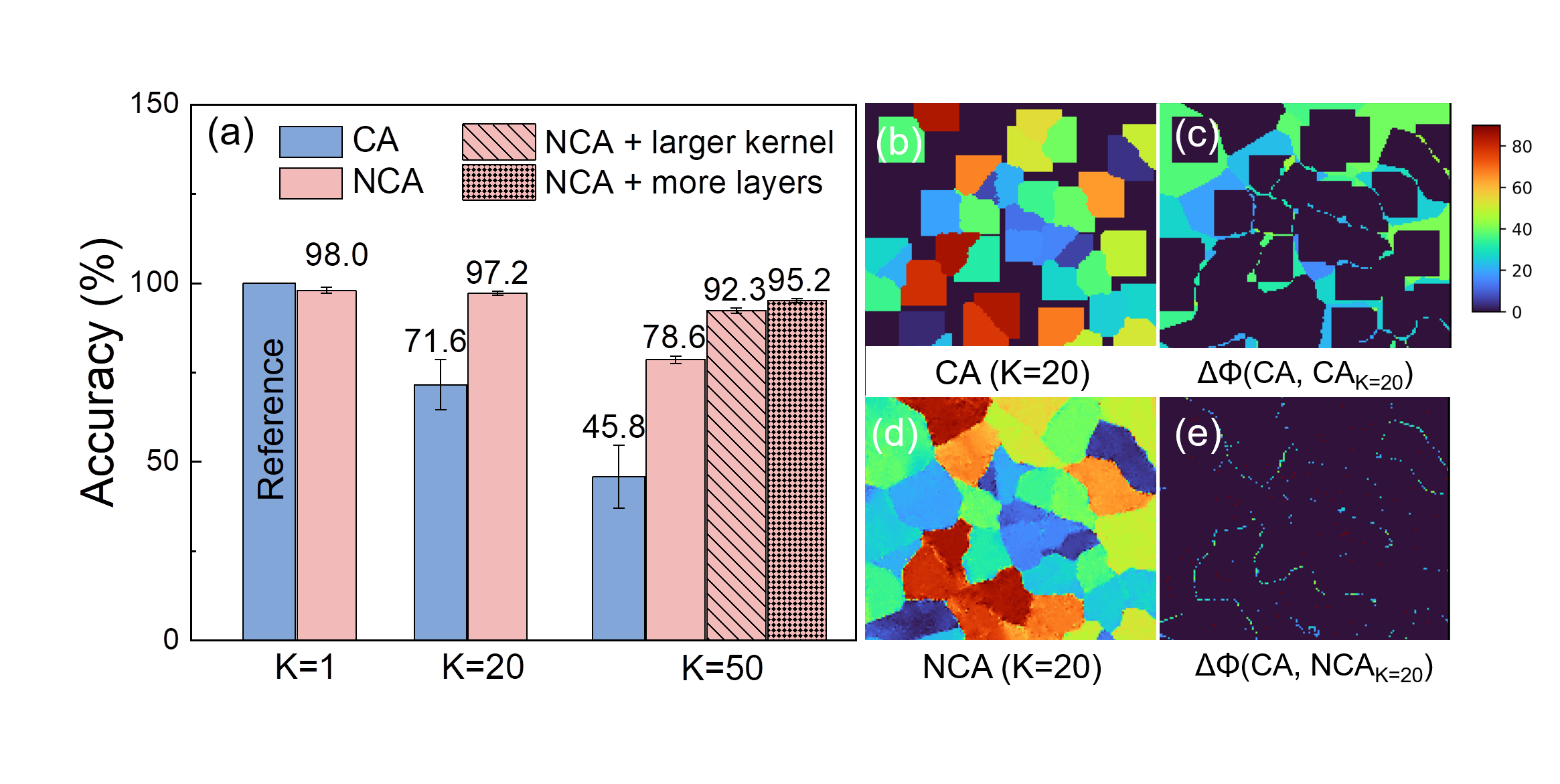}
	\caption{(a) Accuracy of CA and NCA results with different acceleration factors of K=1, 20, 50; microstructure of 2D multi-grain growth with K=20 from CA (b) and NCA (c); orientation difference $\Delta\phi$ of CA  (c) and NCA results (e) with K=20 with respect to CA reference (K=1).}
	\label{fig:incre}
\end{figure}}

\section{Conclusions}\label{sec:conclusion}
In this work, we have presented NCA, a deep-learning-based approach for modelling solidification microstructures. NCA leverages a multi-layer CNN to replace the simple rules of CA, providing higher computational efficiency and flexibility. The convolution operator in NCA is particularly relevant to the physics of solidification, which is a time-progressive phenomenon governed by temperature and microstructure state in the vicinity of the solidification interface. Importantly, NCA allows for encoding physical knowledge into the framework by adopting a 'physics-based activation function', ensuring the satisfaction of solidification physics while easing the training process and reducing the required data size.

We have demonstrated that NCA effectively learns from a relatively small number of CA simulations, since each CA simulation generates a very large dataset by exploring the spatiotemporal evolution of microstructure states for each cell. Interestingly, the trained NCA is independent of boundaries/initial conditions and can accurately predict solidification microstructures even for unseen scenarios during training. This is evident from the high test accuracy when testing the models for larger domain sizes, longer solidification durations, various initial nuclei settings and diverse temperature fields. Moreover, NCA is proven to successfully reproduce the solidification with continuous nucleation (see \textbf{\ref{sec:nuc}}), where nuclei are activated at various time steps, while all the nuclei in the training data are activated at the initial step.

NCA is shown to maintain a high level of accuracy and is up to six orders of magnitude faster than conventional CA, thanks to the efficiency of the adopted computational algorithm, compatibility with GPU computing, and ability to adopt larger time increments for simulations.

Several future extensions of the NCA framework are worth exploring. Firstly, NCA can be trained by data from other high-fidelity microstructure simulations, such as those based on phase-field models, to tackle more complex problems, e.g.\ the solute atom distribution during phase transition. Training such an NCA model can be achieved by introducing solute concentration as additional input and output channels and implementing a physics-informed loss constraint based on solute atom mass conservation.

Secondly, optimizing the training data and CNN architecture can further improve the predictability and efficiency of NCA. For instance, incorporating larger domain sizes in the training data or using a CNN with larger neighbourhood sizes can capture grain growth under larger undercoolings or time increments, as demonstrated in \textbf{\ref{sec:largeT}} and \textbf{Section \ref{sec:computation}}. Adopting a 3D-CNN in NCA can extend the current model into 3D microstructure simulations. Finally, further computational acceleration might be achieved by replacing the CNN with a Fourier Neural Operator (FNO) \cite{FNO}, which employs the Fast Fourier Transform to compute convolutions more efficiently. In summary, we believe NCA bears great potential as a universal method for fast and accurate microstructure simulation.

\section*{Declaration of Competing Interest}\label{sec:comint}
The authors declare that they have no known competing financial interests or personal relationships that could have appeared to influence the work reported in this paper.

\section*{Acknowledgments}\label{sec:Acknowledgements}
Financial support by the Swiss National Science Foundation (SNSF; grant number 200551) is gratefully acknowledged.

\section*{Data availability}\label{sec:Supplementary}
The models and scripts used in this study are available on \href{https://github.com/HighTempIntegrity/JianTang-NCA01}{Github}.

%\section*{Author contributions}\label{sec:aurthor}
%The project was conceptualized and led by E.H. and L.D.L., who also supervised the work. J.T. was responsible for the CA simulations and training the NCA models. All authors contributed to the analysis of results and the writing of the paper.

\section*{Appendices}
\appendix
The NCA simulation strategy is developed to capture the physics of the solidification microstructure formation, in which a CNN is trained by the outcomes of CA simulations. The details of the CA method and the generated data for training and validation of NCA are presented in \textbf{\ref{sec:traindata}}. \textbf{\ref{sec:dgv}} explains the derivation of the dendrite growth rate of Hastelloy X for consideration in the CA simulations. \textbf{\ref{sec:nuc}} and \textbf{\ref{sec:largeT}}  demonstrate the applicability of NCA for simulating consecutive nuclei activation and solidification with larger undercooling ranges. Finally, \textbf{\ref{sec:postpro}} elaborates a post-processing method for NCA and its effect on the predicted microstructure.

\section{CA simulations for training \& validation data}\label{sec:traindata}

CA simulations are used to generate data for training, validation and ultimately testing NCA. \textbf{Algorithm~\ref{alg:ca_algor}} describes the pseudocode for the performed CA simulation in this study. As mentioned in \textbf{Section \ref{sec:architecture}} of the main text, the CA domain is discretized into cells, each containing information such as cell state $P$, Euler angles $\theta$, temperature $T$, etc. The incremental evolution of the information of the cell at every increment $\delta t$ is governed by the information of the cells within a 3$\times$3 neighbourhood in the previous increment.

At the beginning of the simulation, the nuclei at locations $nuc\_{pos}$ are activated in a $nx\times ny$ liquid domain by assigning them the cell state of 'growing' and dedicated Euler angles $nuc\_{ea}$ \cite{lian2019cellular}.  Based on the 'decentred octahedron method' developed by Gandin and Rappaz \cite{gandin19973d}, a growing octahedron with the size of 0 is considered at the centre of the nuclei cells whose diagonals make angles as $nuc\_{ea}$ with the reference coordinate system. The octahedron centre location $Oct$ and size $\lambda$ are tracked during the CA simulations. The incremental growth of the nuclei is simulated through expanding diagonals of the octahedra at a rate given by the 'dendrite growth velocity'.  The dendrite growth velocity is calculated based on the cell undercooling (i.e. $\Delta T=T_{melting}-T$) at each increment and as described in \textbf{\ref{sec:dgv}}. 
\begin{figure}[h!]
	\centering
	\renewcommand{\thefigure}{\arabic{figure}}
	\includegraphics[width=0.45\linewidth]{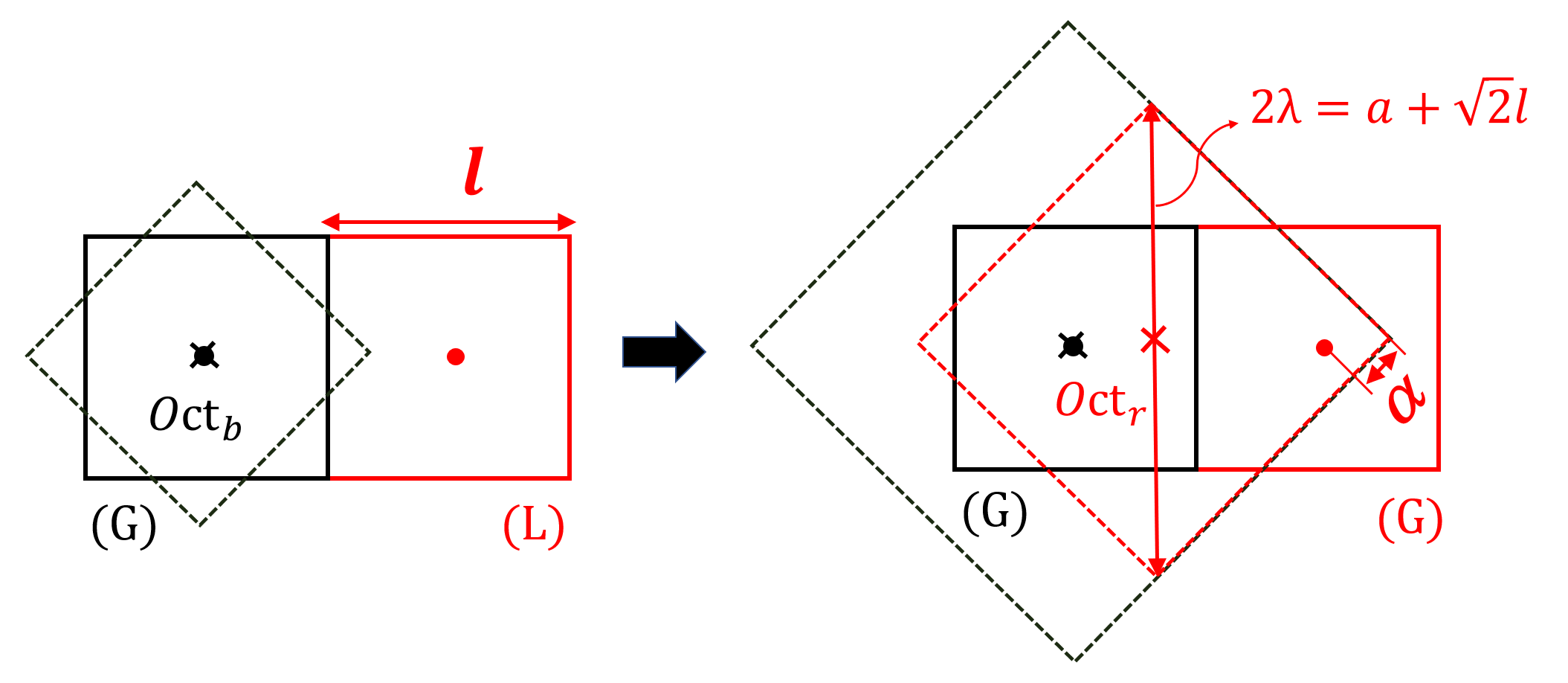}
	\caption{Schematic of the 'decentred octahedron method' for grain growth in CA simulations.}
	\label{fig:oct}
\end{figure}

\textbf{Figure \ref{fig:oct}} illustrates a schematic of the decentred octahedron method for representing the movement of the solidification interface. For each increment, it is mathematically checked if the centre of the liquid cells in the neighbourhood of each growing cell falls inside the octahedron and if so, the state of the liquid cell is changed to growing, and a new octahedron is assigned for it.  The new octahedron shares one corner with the previous octahedron, hence has the same spatial orientation (i.e.\ the Euler angles of the new growing cell are inherited from the parent cell). The size of the new octahedron is derived based on the cell length $l$ and the distance $a$ between the cell centre and the corner of the octahedra, see \textbf{Figure \ref{fig:oct}}.  The new octahedron then grows independently of the parent and based on its local undercooling.  

The transition of the state growing to solid occurs when the growing cell has no neighbour in the liquid state.  When there is no liquid cell in the domain, the simulation is finished. Our developed Python CA code can be accessed from \href{https://github.com/HighTempIntegrity/JianTang-NCA01}{Github} and readers are referred to Ref. \cite{lian2019cellular} for more details about the CA solidification microstructure modelling.

Various CA simulations were performed for generating training, validation and test data of the NCA.  Except for the single-grain growth study (Section \ref{sec:2Dsingle}), the training of the NCA uses CA simulation data from both single-nucleus and multi-nuclei growth. It is observed that the accuracy of NCA in predicting multi-grain growth is improved, when the training data includes both single-nucleus and multi-nuclei CA simulations, especially for cases with a time increment factor K$>$1. For example, when K=5, the NCA model trained with the dataset in \textbf{Section \ref{sec:2Dmulti}} obtains a better test accuracy of 96.8$\pm$0.8\% compared with the model trained by the same amount of data with only multi-nuclei CA simulations (test accuracy of 95.3$\pm$0.9\%). However, the performance of the NCA model trained with only multi-nuclei data can be improved if the training data size increases, e.g.\ the NCA model trained with 80 multi-nuclei CA simulations obtains a test accuracy of 97.7$\pm$0.5\% for K=5.

\begin{algorithm}[hbt!]
\normalsize	
\caption{Pseudocode of the CA simulation for generating training, validation, and testing data}\label{alg:ca_algor}
\begin{algorithmic}[1]
\Require{$nx$, $ny$, $T(x,y,t)$, $nuc\_{pos}$, $nuc\_{ea}$, $\delta$t}
\Ensure{$P$, $\theta$}
 \State \texttt{$P \gets -1, \theta \gets (0, 0, 0), t\gets0$}; \Comment{initialize a nx$\times$ny liquid domain}
 \State \texttt{$P[nuc\_pos] \gets 0$, $\theta[nuc\_pos] \gets nuc\_ea$}; \Comment{activate the nuclei at the initial step}
 \State  \texttt{Half length of octahedron diagonal $\lambda[nuc\_pos] \gets 0$}; \Comment{initialize octahedron}
 \State  \texttt{Octahedron centre $Oct[nuc\_pos] \gets Pos[nuc\_pos]$};
 \State \texttt{Count the number of liquid cells ($P=-1$) $Liq\_num$};
 \While{\texttt{$Liq\_num \neq$ 0}}
  \State \texttt{find the growing cell $C_G$};
  \State \texttt{update octahedron size $\lambda[C_G] \gets \lambda[C_G] + v(T[C_G]-T_m) \times \delta t $};
  \State \texttt{------------------- Grain growth -------------------}
  \State \texttt{find the liquid cells with neighbour growing cells $C_{Ln}$};
  \For{\texttt{each $C_{Ln}$}}
    \State \texttt{$d_{small}=0$};
    \For{\texttt{each of their growing neighbour cells $C_{Gn}$}}
        \State \texttt{$Vec=Pos[C_{Ln}] - Oct[C_{Gn}]$};
        \State \texttt{$Vec'=Rot(\theta[C_{Gn}])^{-1}\times Vec$};
        \State \texttt{$d_i = \lambda[C_{Gn}] - sum(|Vec'|)$};
        \If{ \texttt{$d_i \leq d_{small}$}}
            \State \texttt{$d_{small} \gets d_i$};
            \State \texttt{$C_{cap} \gets C_{Gn}$};
        \EndIf
    \EndFor
    \If{ \texttt{$d_{small} < 0$}} 
            \State \texttt{$P[C_{Ln}] \gets 0$, $\theta[C_{Ln}] \gets \theta[C_{cap}]$};
            \State \texttt{assign a new octahedron for $C_{Ln}$}; 
    \EndIf    
  \EndFor
 \State \texttt{---------------------------------------------------}
   \For{\texttt{each $C_{G}$}}
    \If{ \texttt{no liquid cell in its neighborhood}};
            \State \texttt{$P[C_{G}] \gets 1$};
    \EndIf  
  \EndFor
 \State \texttt{$t \gets t+\delta t$}
 \State \texttt{Count the number of liquid cells $Liq\_num$};
 \EndWhile
\end{algorithmic}
\end{algorithm}

\newpage

\section{Dendrite growth velocity}\label{sec:dgv}
Although there are examples of experimental measurement of dendrite growth rate, e.g.\ in \cite{dgv1}, this is often estimated from empirical models \cite{lipton1984dendritic,kurz1986theory,kurz1989fundamental} or phase-field simulations \cite{dgv2}. One of the prevailing empirical models, the Kurz-Giovanola-Trivedi (KGT) model \cite{kurz1986theory,kurz1989fundamental} describes the relation between the dendrite tip radius $R$ and its growth velocity $v$ by the following two equations:

\begin{equation}\label{eq:6}
P_c = \frac{vR}{2D}
\end{equation}
\begin{equation}\label{eq:7}
G_c(P_c,v)\epsilon_c(P_c)= \frac{4\pi^2\Gamma}{m_lR^{2}} + \frac{G}{m_l}
\end{equation}
where $G$, $G_c$, $P_c$, $D$, $m_l$, $\Gamma$ are the mean thermal gradient at the dendrite tip, the concentration gradient in the liquid near the dendrite tip, the Peclet number, the liquid diffusivity, the liquidus slope, and the Gibbs-Thomson coefficient, respectively. The thermal gradient $G$ is neglected in the dendritic growth regime \cite{rappaz1993probabilistic}. The concentration gradient $G_c$ and the dimensionless variable $\epsilon_c$ are functions of the Peclet number $P_c$ and growth rate $v$ \cite{Rappaz1989,kurz1986theory}:

\begin{equation}\label{eq:8}
G_c(P_c,v) = \frac{c_0(1-k)v/D}{I_v(P_c)(1-k)-1}
\end{equation}

\begin{equation}\label{eq:9}
I_v(x) = x\cdot exp(x)\int\frac{exp(-x)}{x} dx
\end{equation}

\begin{equation}\label{eq:10}
\epsilon_c(P_c) = 1-\frac{2k}{\left[1+(2\pi/P_c)^2\right]^{0.5}+2k-1}
\end{equation} 
where $k$, $c_0$, and $I_v(x)$ are the equilibrium partition coefficient, the solute atom concentration, and the Ivantsov function, respectively. 
Often the constitutional undercooling, resulting from the solute atom segregation, is considered as the main contributor to the supercooling \cite{lipton1984dendritic,kurz1986theory,kurz1989fundamental}.  The constitutional undercooling with a given Peclet number $P_c$ is:

\begin{equation}\label{eq:11}
\Delta T = m_lc_0\left[1-\frac{1}{1-(1-k)I_v(P_c) }\right]
\end{equation}

The above set of equations can be solved for a range of $v\cdot R$ values and accordingly the dependence of the dendrite growth rate $v$ on the undercooling $\Delta T$ can be derived.  For the sake of simplicity, a polynomial is ultimately fitted to describe this dependence, i.e.\ $v(\Delta T)$. 

For the application of the KGT model to multi-component systems, the diffusion field of each solute species should be superimposed, and hence, the undercooling $\Delta T$ and dendrite tip radius $R$ are \cite{dgv6,kurz1986theory,kurz1989fundamental}:

\begin{equation}\label{eq:12}
\Delta T = \sum_i^N {\Delta T_i}
\end{equation}
\begin{equation}\label{eq:13}
R = \left[\frac{4\pi^2\Gamma}{ \sum_i^N {m_l^iG_c^i(P^i_c,v)\epsilon^i_c(P^i_c)}-G}\right]^{0.5}
\end{equation}
\begin{equation}\label{eq:14}
P^i_c = \frac{vR}{2D_i}
\end{equation}
We assume the Hastelloy X as a Ni-20Cr-20Fe-10Mo alloy and use the parameters summarised in \textbf{Table \ref{tab:materials}} to derive \textbf{Equation \ref{eq:1}} (in the main text) for its dendrite growth rate.

\mbox{}
\newline

\begin{table}[h!]
\centering
\renewcommand{\thetable}{\arabic{table}}
\caption{Material properties of Hastelloy X alloy for deriving its dendrite growth velocity \cite{hxp1,hxp2,hxp3,hxp4}.}\label{tab:materials}
\begin{tabular}{llll}
\toprule
System &	Ni-Cr & Ni-Fe & Ni-Mo\\
\midrule
Liquidus slope $m_l$ $(K\cdot wt\%^{-1})$& -3.166& -1.261& -2.628\\
Solute composition $c_0$ $(wt\%$)&20& 20 &10\\
Equilibrium partition coefficient $k$& 0.594 &0.986 &0.99\\
Liquid diffusivity $D$ $(10^{-9}\cdot m^2s^{-1})$& 3.04 &3.20 &8.861\\
Gibbs-Thomson coefficient $\Gamma$ $(K\cdot m)$& & $3 \times 10^{-7}$&\\
\bottomrule
\end{tabular}
\end{table}

\section{NCA for consecutive nuclei activation}\label{sec:nuc}

\begin{figure}[!h]
	\centering
	\renewcommand{\thefigure}{\arabic{figure}}
	\includegraphics[width=0.45\linewidth]{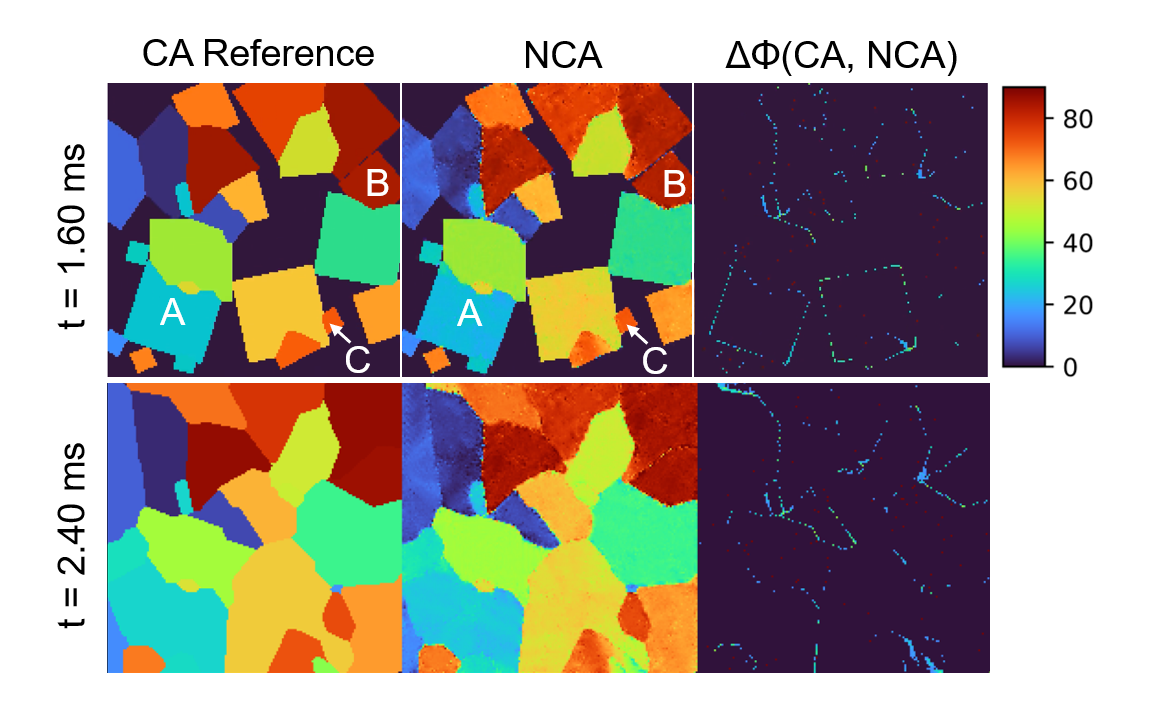}
	\caption{Test case of 2D multi-grain growth with consecutive nucleation (160$\times160 \; \micro$m$^2$ domain with nucleation density of 1562 mm$^{-2}$). Grains A, B and C are activated consecutively during the solidification process.}
	\label{fig:nucact}
\end{figure}

The results reported in the main text discuss the application of the NCA for conditions in which all nuclei are activated at the start of the simulation. However, solidification microstructure modelling for processes such as MAM requires consideration of continuous nucleation.  We examined the ability of the trained NCA in \textbf{Section \ref{sec:2Dmulti}} to predict microstructure development for conditions where nuclei are activated consecutively during the solidification process.  \textbf{Figure~\ref{fig:nucact}} shows the NCA model predictions in comparison with CA results, showing that the trained NCA can well represent the microstructure evolution with consecutive nucleation. 

\begin{figure}[!h]
	\centering
	\renewcommand{\thefigure}{\arabic{figure}}
	\includegraphics[width=0.45\linewidth]{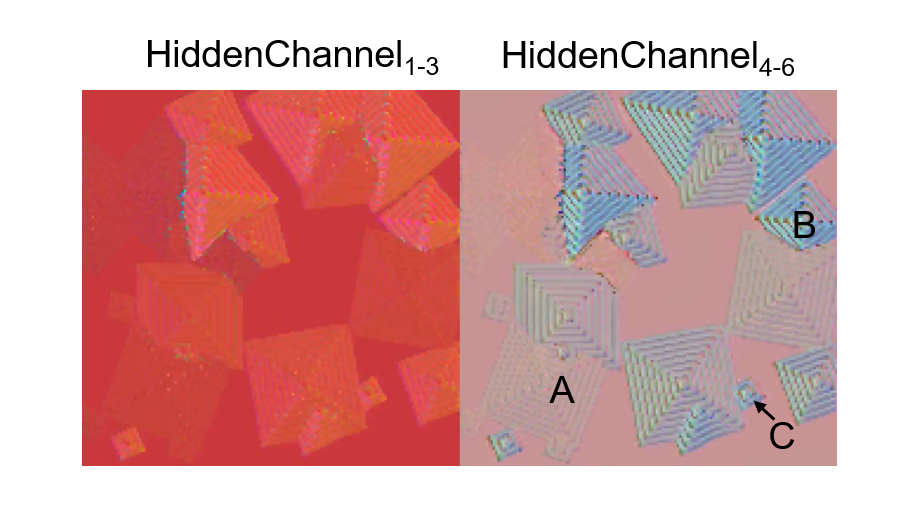}
	\caption{Hidden channels of the test case in Figure~\ref{fig:nucact}. The RGB values of the pixel colours are the values in the hidden channels normalized by their maximum and minimum values, i.e.\ $H^i_{normalized} = (H^i-H^i_{min})/(H^i_{max}-H^i_{min})$.}
	\label{fig:nucact_hid}
\end{figure}
\textbf{Figure~\ref{fig:nucact_hid}} represents the values of hidden channels 1-3 and 4-6 for the simulation described above.  The RGB values of the pixel colours are the values in the hidden channels normalized by their maximum and minimum, i.e.\ $H^i_{normalized} = (H^i-H^i_{min})/(H^i_{max}-H^i_{min})$. Interestingly, an octahedron pattern is visible for each grain, where the size and orientation of the diagonals are related to the grain size and orientation, i.e.\ similar to the concept of decentred octahedron discussed in \textbf{\ref{sec:traindata}}.

\section{NCA for larger undercoolings}\label{sec:largeT}
\begin{figure}[!h]
	\centering
	\renewcommand{\thefigure}{\arabic{figure}}
	\includegraphics[width=0.45\linewidth]{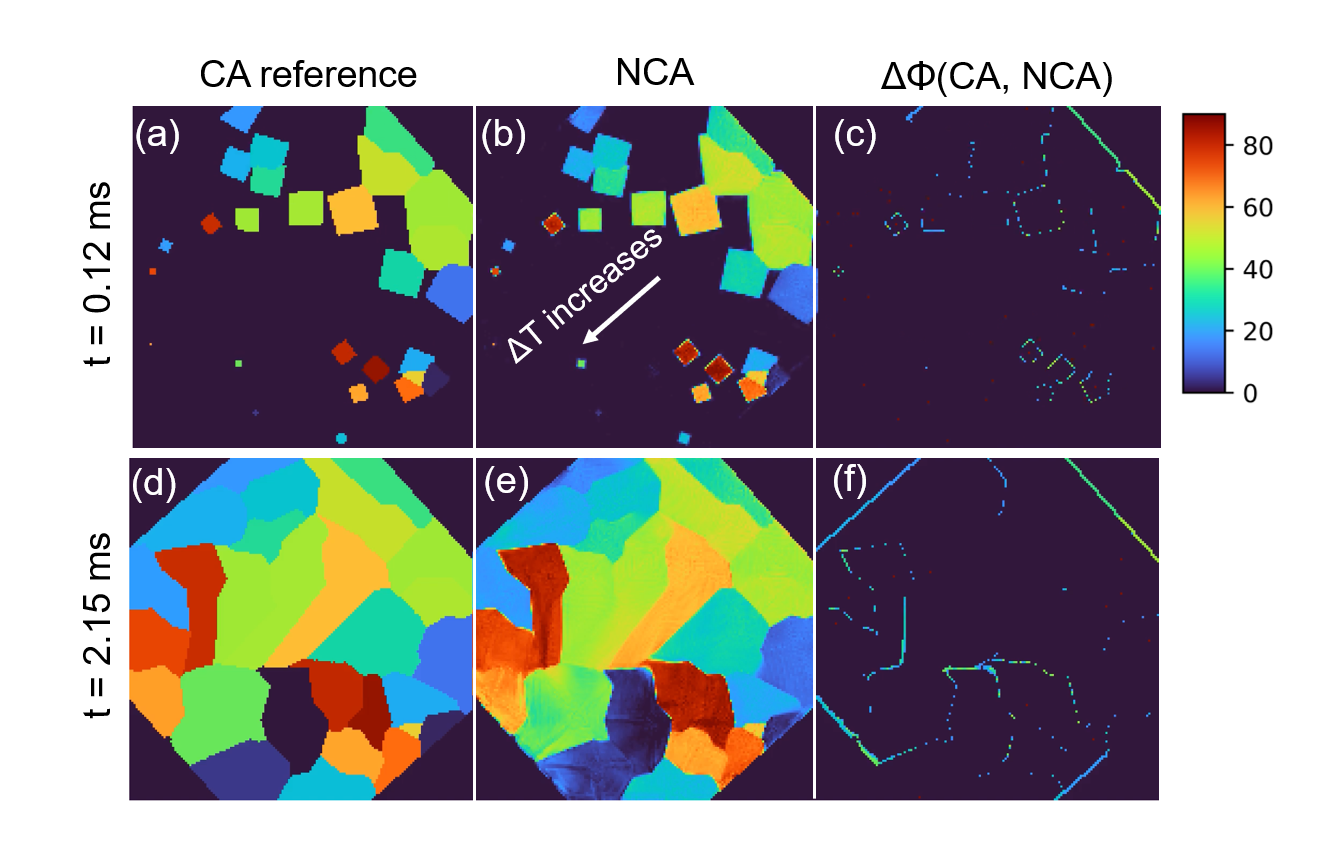}
	\caption{Test case for 2D multi-grain growth with a random temperature gradient (160$\times160 \; \micro$m$^2$ domain with nucleation density of 1562 mm$^{-2}$): CA results (a and d), NCA predictions (b and e), and orientation difference $\Delta\phi$ between them (c and f). A temperature gradient with maximum and minimum undercooling of 15 and 45 K is applied as shown in (b).}
	\label{fig:temprot}
\end{figure}

\begin{figure}[!h]
	\centering
	\renewcommand{\thefigure}{\arabic{figure}}
	\includegraphics[width=0.45\linewidth]{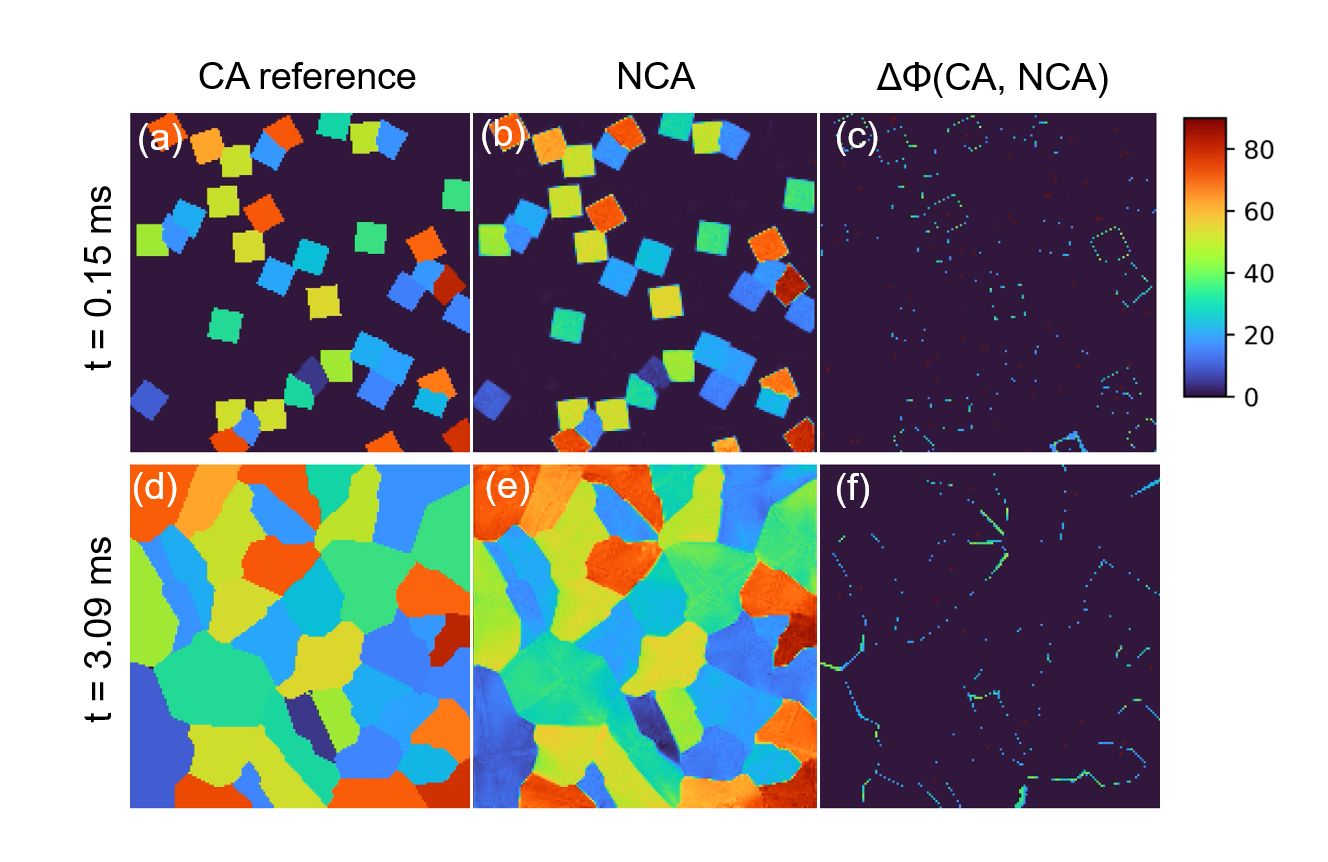}
	\caption{Test case of 2D multi-grain growth under continuous cooling with an initial undercooling of 15 K and cooling rate of $1.5\times10^4 \;$ K/s (160$\times160 \; \micro$m$^2$ domain with nucleation density of 1562 mm$^{-2}$): CA results (a and d), NCA results (b and e), and orientation difference $\Delta\phi$ between them (c and f). Note that the NCA did not see any data for continuous cooling during training.}
	\label{fig:tempcool}
\end{figure}

The main text describes training and validation of NCA for up to 25 K undercooling. Application of the NCA for simulating fast solidification during MAM requires consideration of larger undercoolings. This section examines the effectiveness of NCA for solidification microstructure simulation with undercooling up to 45 K.  This involves consideration of CA simulations from a larger domain size of 128$\times$128 $\micro$m$^2$ for training NCA. The CA simulates solidification under a temperature gradient with undercooling in the range of 15 and 45 K and for nuclei density of 1831-3662 mm$^{-2}$ to generate data for training and validation of NCA. NCA are trained with a fast decaying learning rate (decaying 70\% every 100 steps) for 250 epochs. The model obtains accuracies of 97.7\%, 97.4\%, and 96.2\% for training, validation and testing, respectively, see \textbf{Figure~\ref{fig:temprot}}.  The trained model can also simulate solidification under continuous cooling for the mentioned undercooling range, as presented in  \textbf{Figure~\ \ref{fig:tempcool}}.

%\newpage

\section{Post-processing of NCA results}\label{sec:postpro}
Here describes a post-processing technique to remove the artefacts in the NCA predictions, e.g.\ the fluctuations of Euler angles within the grains (see \textbf{Section \ref{sec:2Dmulti}}). As the orientation of grains inherits from its nuclei, the main idea of this post-processing method is to replace the pixel Euler angles with that from one of the nearby nuclei that has the most similar orientations. 
For a pixel located at $(x_i, y_i)$  with Euler angles $\boldsymbol{\theta}_i$ and a nucleus $n_j$ in its vicinity, we define $D_{ij}$ as:

\begin{equation}\label{eq:15}
{D}_{ij}= \frac{ \|\boldsymbol{\theta}_{i} - \boldsymbol{\theta}_{n_j}\|^2 } {\max \limits_{1\leq n_k \leq n_{tot}}\{ \|\boldsymbol{\theta}_{i} - \boldsymbol{\theta}_{n_k}\|^2\} }+ \lambda \frac{(x_i-x_{n_j})^2+(y_i-j_{n_j})^2}{\max \limits_{1\leq nl \leq n_{tot}}\{(x_i-x_{n_l})^2+(y_i-j_{n_l})^2\}} 
\end{equation}

\begin{figure}[!h]
	\centering
    \renewcommand{\thefigure}{\arabic{figure}}
	\includegraphics[width=0.55\linewidth]{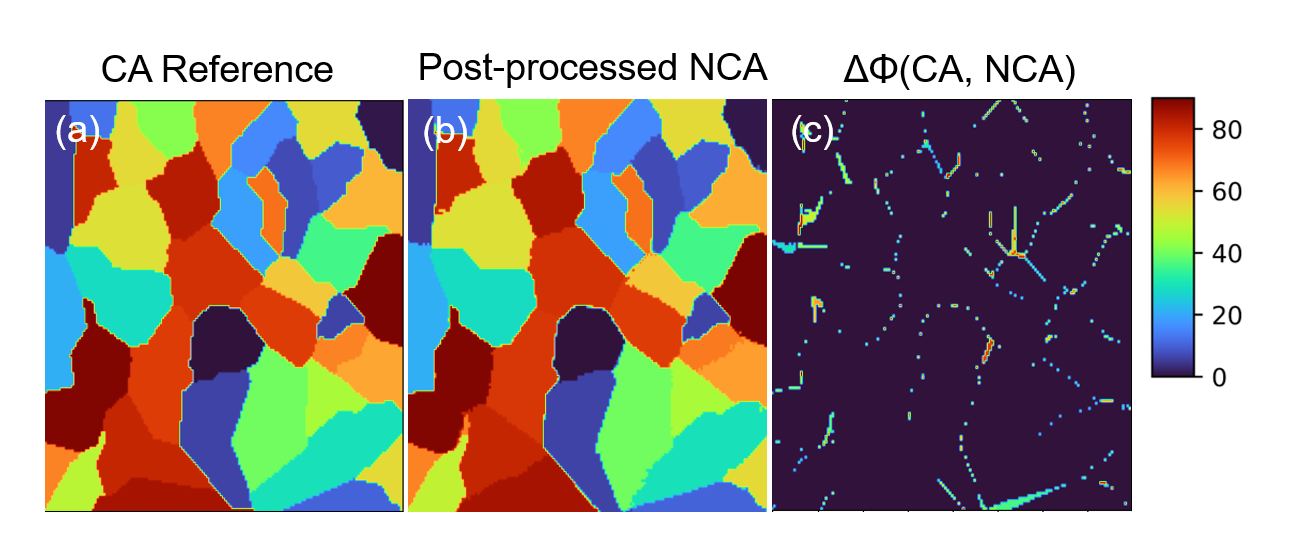}
	\caption{Comparison of CA and post-processed NCA results from \textbf{Figure \ref{fig:2Dmulti}}: (a) the CA result, (b) the NCA prediction after post-processing, (c) the orientation difference between them.}
	\label{fig:postpro}
\end{figure}

$D_{ij}$ is an index representing the differences of Euler angles and location between pixels and the nucleus. In the post-processing of NCA results, we replace the pixel's Euler angles with that of the nuclei with the smallest ${D}^{ni}_{i,j}$. \textbf{Figure \ref{fig:postpro}} shows a comparison between the CA and the post-processed NCA results from \textbf{Figure \ref{fig:2Dmulti}}. Most of the artefacts are removed after post-processing with only a few inconsistencies at the grain boundaries compared with CA results.

\bibliographystyle{unsrt}
\biboptions{sort&compress} 
%\bibliographystyle{abbrvnat}
%\bibliographystyle{unsrt}
%\biboptions{sort&compress}
\bibliography{r_refs}

\end{document}